\documentclass[conference,compsoc]{IEEEtran}

\usepackage{amsmath,amssymb,amsfonts}
\usepackage{graphicx}
\usepackage{float}
\usepackage{booktabs}
\usepackage{multirow}
\usepackage{xcolor}
\usepackage{url}
\usepackage[hidelinks]{hyperref}
\usepackage{bm}
\hypersetup{
  pdftitle={Toward Interpretable Privacy Guarantees in Face-Swapping Anonymization},
  pdfauthor={Vishnu Bondalakunta, Arman Zareian Jahromi, Shuangqing Wei, George Amariucai},
  pdfkeywords={face-swapping, face anonymization, identity leakage, membership inference}
}

\newcommand{\R}{\mathbb{R}}
\newcommand{\D}{\mathcal{D}}
\newcommand{\Ssub}{\mathcal{S}}
\newcommand{\G}{G}
\newcommand{\F}{\mathcal{F}}
\newcommand{\FG}{\mathcal{F}_G}
\newcommand{\Amat}{\bm{A}}
\newcommand{\Bmat}{\bm{B}}
\newcommand{\cvec}{\bm{c}}
\newcommand{\evec}{\bm{e}}
\newcommand{\norm}[1]{\left\lVert#1\right\rVert}
\newcommand{\ind}[1]{\mathbb{I}\!\left[#1\right]}
\newtheorem{prediction}{Prediction}

\newcommand{\npairs}{947}

\newcommand{\ffTrain}{57{,}048}
\newcommand{\ffTest}{10{,}667}
\newcommand{\fitcos}{0.769}
\newcommand{\fitRtwo}{0.585}
\newcommand{\stochfrac}{0.414}
\newcommand{\rhoB}{0.894}
\newcommand{\sigmaxB}{10.1}

\newcommand{\buGain}{0.183}
\newcommand{\rhoAval}{1.12}
\newcommand{\modelresidmed}{14.9}
\newcommand{\intraidmed}{13.7}
\newcommand{\rhoStableN}{5{,}000}
\newcommand{\fitStableN}{20{,}000}

\newcommand{\bfTrain}{56{,}848}
\newcommand{\bfTest}{10{,}397}
\newcommand{\bfFitCos}{0.647}
\newcommand{\bfRtwo}{0.399}
\newcommand{\bfRhoB}{0.920}
\newcommand{\bfSigma}{15.1}
\newcommand{\bfBu}{0.219}
\newcommand{\csTrain}{16{,}849}
\newcommand{\csTest}{7{,}239}
\newcommand{\csFitCos}{0.599}
\newcommand{\csRtwo}{0.338}
\newcommand{\csRhoB}{0.854}
\newcommand{\csSigma}{7.81}
\newcommand{\csBu}{0.198}

\newcommand{\Nchains}{478}
\newcommand{\Kpass}{5}
\newcommand{\selfsim}{0.616}

\newcommand{\mleakfive}{0.047}
\newcommand{\mfloor}{0.006}
\newcommand{\mratioearly}{0.139}
\newcommand{\mratiolate}{0.893}
\newcommand{\maucone}{0.874}
\newcommand{\mauclo}{0.709}
\newcommand{\passestofloor}{18\text{--}20}

\begin{document}

\title{Toward Interpretable Privacy Guarantees in Face-Swapping Anonymization}

\author{\IEEEauthorblockN{Vishnu Bondalakunta\IEEEauthorrefmark{1},
Arman Zareian Jahromi\IEEEauthorrefmark{1},
Shuangqing Wei\IEEEauthorrefmark{2}, and
George Amariucai\IEEEauthorrefmark{1}}
\IEEEauthorblockA{\IEEEauthorrefmark{1}Kansas State University, Manhattan, KS, USA\\
\IEEEauthorrefmark{2}Louisiana State University, Baton Rouge, LA, USA}}

\maketitle

\begin{abstract}
Face-swapping has emerged as a promising approach to facial privacy protection, replacing a target
individual's appearance with that of a donor while preserving non-facial context. The resulting
images visually resemble the donor, and face recognition systems tend to suppress the target's
match scores --- ostensibly satisfying privacy requirements. Empirical evaluation across a range of
face-swapping models, however, reveals that significant target identity leakage still occurs. This
raises a deeper question: why does leakage occur, and can it be predicted? We propose a linear
stochastic model that treats face-swappers as transformations on the space of identity embeddings,
providing an interpretable account of the leakage mechanism. The model is fit to empirical
observations and used to derive testable predictions. The aim is to ground privacy assessments in
principled, interpretable analysis, thus making formal privacy guarantees explainable --- and
perfectible --- rather than purely observational.
\end{abstract}

\begin{IEEEkeywords}
face-swapping, face anonymization, identity leakage, membership inference, linear stochastic
systems, face embeddings.
\end{IEEEkeywords}

\section{Introduction}
\label{sec:intro}
\IEEEPARstart{F}{acial} imagery is collected at an enormous scale: hospitals archive clinical
photographs, researchers assemble vision datasets, municipalities operate street cameras, and
companies log video for quality assurance. Much of this imagery is valuable precisely because it
shows people --- their expressions, their poses, their interactions with an environment --- yet the
faces it contains are biometric identifiers. Releasing such data without protection exposes the
photographed individuals to re-identification, and modern data-protection regulation such as HIPAA ~\cite{hipaa1996} and GDPR ~\cite{gdpr2016} increasingly treats facial images as sensitive personal data. The data owner therefore faces a familiar
dilemma: discard the imagery, destroy its utility through blurring or pixelation, or find an
anonymization mechanism that removes identity while preserving everything else.

Face-swapping is an attractive candidate for the third option. A face-swapping model takes two
images --- a \emph{donor}, whose identity may be used freely, and a \emph{target}, whose identity
must be protected --- and synthesizes an output carrying the donor's facial identity in the
target's pose, expression, and scene context. To a human observer the output depicts the donor;
to a face recognition system the target's match score drops sharply; and unlike blurring, the
output remains a photorealistic face that downstream applications (expression analysis, gaze
estimation, crowd analytics, dataset publication) can still
consume. A growing body of work consequently proposes face-swapping, or closely related face
synthesis, as a privacy
mechanism~\cite{gafni2019live,maximov2020ciagan,hukkelas2023deepprivacy2,wen2022identitydp,barattin2023attribute}.

The implicit security argument runs as follows: the swap looks like the donor; face recognition
says it matches the donor; therefore the target is gone. The first part of this paper subjects
that argument to a direct adversarial test and finds it wanting. Across seven publicly available
face-swapping systems spanning GAN-based, encoder-decoder, and diffusion-based architectures, we
generate roughly one thousand swaps per tool under a controlled protocol and measure how much
\emph{target} identity survives in the output, as seen by independent face recognition embeddings.
The result is unambiguous: every tool we test leaks the target. Swap outputs remain measurably and
exploitably closer to other photographs of the target than to photographs of unrelated people:
an adversary who merely thresholds an off-the-shelf similarity score can, for several popular
tools, identify a substantial fraction of protected individuals while raising almost no false
alarms. Yet for five of the seven tools the output is strongly donor-dominated and would pass any
visual or donor-centric audit; the leakage is invisible unless one explicitly looks for it.

Establishing leakage, however, only sharpens the real question. A purely empirical evaluation
reports that a tool leaks a given amount, but not \emph{why}; it cannot extrapolate beyond the
operating points it measured or guide interventions.
Consider the most natural counter-measure: if one swap suppresses the target signal by an order of
magnitude, will a second swap --- re-swapping the output with a fresh donor --- suppress it by another
order of magnitude? Will repeated swapping drive the leakage to zero, or to a floor? How many
passes are enough? These are questions about the \emph{mechanism} of leakage, and answering them
requires a model.

The second and central part of this paper proposes such a model. We treat the face-swapper not as
an image-to-image black box but as an operator acting on the identity-embedding space of the face
recognition system used by the adversary. The modeling step is deliberately simple: we approximate
the swap embedding as an affine function of the donor and target embeddings plus a stochastic
residual,
\[
  \bm{s} \;=\; \Amat\bm{d} + \Bmat\bm{t} + \cvec + \evec ,
\]
where $\bm{d}$, $\bm{t}$, and $\bm{s}$ are the donor, target, and swap embeddings, $\Amat$ and
$\Bmat$ are fixed matrices, $\cvec$ is a bias vector, and $\evec$ is a zero-mean noise term.
We do not claim the image generator is linear --- it manifestly is not --- but we show this
first-order approximation in embedding space is accurate enough to be useful: fitted by ridge
regression on identity-disjoint corpora of tens of thousands of swaps, it explains
$34$--$59\%$ of held-out swap-embedding variance for the three tools we model (FaceFusion,
BlendFace, CanonSwap), substantially outperforms donor-copy and donor-only baselines, and leaves
a residual comparable in scale to natural within-person embedding variation
and --- crucially --- nearly uncorrelated with the target identity. Once this approximation is
accepted, repeated swapping becomes a linear stochastic dynamical system: the
target signal entering a cascade of swaps is propagated only through powers $\Bmat^k$, so its
fate is governed by the spectral characteristics of the fitted target-transfer operator $\Bmat$.

The model makes quantitative, falsifiable predictions: leakage under repeated swapping should fade
at two distinct rates (a fast first-pass contraction set by the Rayleigh gain of $\Bmat$ along
real target directions, and a slow asymptotic tail set by the spectral radius $\rho(\Bmat)$);
the leakage floor should equal the \emph{non-member} similarity baseline --- the score an unrelated
identity, absent from the release, attains against the swap --- rather than zero; and tools
should be ordered in cascade-tail speed by their fitted spectral radii. We test these predictions
on a generated five-pass FaceFusion cascade of \Nchains{} chains
with strict no-reuse protocols, and three-pass BlendFace and CanonSwap cascades of roughly
one thousand chains per tool. The headline result is that the asymptotic prediction is strikingly
accurate --- the measured late-pass decay ratio ($\mratiolate$) matches the independently fitted
spectral radius ($\rhoB$) to three decimal places --- while the predicted absolute leakage level is
less accurate, in an instructive direction that we analyze honestly: the model over-predicts how
much later passes erase. Dilution helps, but more slowly than naive simulation suggests, and a
membership adversary still achieves AUC $\mauclo$ after five passes.

\smallskip
\noindent\textbf{Contributions.} In summary, this paper makes the following contributions:
\begin{itemize}
  \item \textbf{A systematic leakage study (Part 1).} Under a unified VGGFace2 protocol
  (\npairs{} donor/target/non-member triples per tool), we quantify target leakage for seven
  modern face-swappers via score distributions, membership-inference TPR at low FPR, and
  closed-set CMC curves, with two independent recognizers and a hardened non-member pool. All
  seven tools leak; five transfer donor identity well enough to be credible anonymizers, making
  their residual leakage a genuine, hidden risk.
  \item \textbf{An interpretable model of the leakage mechanism (Part 2).} We model face-swappers
  as affine stochastic operators on identity-embedding space, fit the operators for FaceFusion,
  BlendFace, and CanonSwap on large identity-disjoint corpora, and validate the modeling
  assumptions explicitly --- reporting which hold, hold partially, and fail --- together with the
  corpus sizes required for stable operator identification.
  \item \textbf{Testable predictions and their verification.} From the fitted operators we derive
  spectral predictions for dilution cascades and test them on multi-pass swap data, reporting
  successes (two-rate decay, the $\rho(\Bmat)$-matched tail, the non-member floor, cross-tool
  ordering) and failures (over-predicted erasure levels, weak persistent-direction effects), the
  latter traced to cascade inputs leaving the natural image-embedding manifold.
  \item \textbf{Actionable implications.} Single-swap anonymization should be presumed leaky;
  dilution improves privacy at a measurable, tool-specific exponential rate
  ($\approx \passestofloor$ passes to the noise floor for the tools studied); and the spectral
  radius of the target-transfer operator is a measurable design target for building --- and
  certifying --- better anonymizers.
\end{itemize}

\smallskip
\noindent\textbf{Organization.} Sections~\ref{sec:related}--\ref{sec:threat} cover background,
notation, and the threat model; Section~\ref{sec:part1} presents the empirical leakage study;
Sections~\ref{sec:model}--\ref{sec:verify} introduce and validate the affine stochastic model,
derive cascade predictions, and test them on dilution data;
Sections~\ref{sec:discussion}--\ref{sec:conclusion} discuss implications, limitations, and
conclusions.

\section{Background and Related Work}
\label{sec:related}

\subsection{Face Recognition and Identity Embeddings}
Modern face recognition is built on deep metric learning. A convolutional backbone, typically a
ResNet variant~\cite{he2016deep}, is trained with a margin-based classification loss such as
ArcFace~\cite{deng2019arcface} or a triplet objective such as FaceNet~\cite{schroff2015facenet}
so that images of the same person map to nearby points of a fixed-dimensional representation
space, while images of different people map far apart. The trained network then serves as a
feature extractor: a face image maps to a unit-norm \emph{identity embedding}, and identity
decisions reduce to cosine-similarity comparisons. The representation is deliberately invariant
to pose, illumination, expression, and age, which is exactly why it is the right instrument for
measuring identity leakage: anything
that survives in embedding space is, by construction, identity-relevant signal. In this work we
use the InsightFace \texttt{buffalo\_l} ArcFace model~\cite{insightface,deng2019arcface} as the
primary extractor and Facenet512~\cite{schroff2015facenet}, as implemented in the DeepFace
library~\cite{serengil2020lightface}, as an independent secondary extractor.

\subsection{Face-Swapping}
Face-swapping transfers the identity of a source (donor) face onto a target image while
preserving the target's pose, expression, and scene. Architecturally, published systems fall into
several families. Encoder-decoder systems in the DeepFaceLab lineage~\cite{perov2020deepfacelab}
learn person-specific or universal autoencoders. GAN-based systems --- including
FaceShifter~\cite{li2020faceshifter}, SimSwap~\cite{chen2020simswap},
FSGAN~\cite{nirkin2019fsgan}, and BlendFace~\cite{shu2023blendface} --- inject an identity
embedding of the donor into a generator conditioned on target attributes; BlendFace specifically
re-designs the identity encoder to reduce attribute entanglement. GAN-inversion systems such as
E4S~\cite{liu2023e4s} perform fine-grained regional editing in the latent space of a pretrained
generator. Recent diffusion-based systems --- DiffSwap~\cite{zhao2023diffswap} and
DiffFace~\cite{kim2025diffface} --- formulate swapping as conditional inpainting or guided
denoising. Video-oriented systems such as CanonSwap~\cite{luo2025canonswap} decouple motion from
appearance to keep swaps temporally consistent. Production pipelines such as
FaceFusion~\cite{facefusion} wrap pretrained swap models (we use its
\texttt{hyperswap\_1a\_256} model) with detection, alignment, and blending stages. Notably,
most of these systems use an ArcFace-style recognition embedding \emph{internally} to represent
the donor identity and to score identity transfer during training --- a point that will matter when
we model their action directly on that embedding space.

\subsection{Face Anonymization and De-Identification}
Early de-identification work established that naive obfuscation (blurring, pixelation, masking)
both degrades utility and fails against trained recognizers, and proposed the $k$-Same family of
formal methods, which replace each face with an average over $k$ individuals to bound
re-identification risk~\cite{newton2005preserving,gross2006model}. Generative anonymizers
followed: DeepPrivacy~\cite{hukkelas2019deepprivacy} inpaints faces with a conditional GAN,
CIAGAN~\cite{maximov2020ciagan} conditions generation on a controllable identity label, and a
broad survey of the area is given by Meden et al.~\cite{meden2021privacy}. Face-swapping is
increasingly proposed as an anonymization primitive because, unlike inpainting, it preserves a
natural-looking face with controllable identity and retains downstream utility.

\subsection{Membership Inference and Privacy Evaluation}
Membership inference attacks (MIAs) ask whether a particular record was part of a model's training
set~\cite{shokri2017membership,yeom2018privacy,salem2019mlleaks}. Beyond their direct threat,
MIAs have become the standard empirical instrument for auditing privacy mechanisms. Carlini et
al.~\cite{carlini2022membership} argue that MIAs must be evaluated by their true-positive rate at
very low false-positive rates (TPR at low FPR), because an attack that confidently identifies even
a small fraction of members is far more damaging than one that is marginally better than chance on
average; we adopt this methodology throughout. In our facial-privacy setting the
analogous question is whether a candidate \emph{identity} is represented inside an anonymized
image release; we formalize this as an \emph{identity inference attack} in
Section~\ref{sec:prelim} and instantiate it both as a membership-style binary decision and as a
closed-set re-identification (ranking) problem evaluated with cumulative match characteristic
(CMC) curves~\cite{phillips2000feret,zheng2016person}.

\subsection{Linear Systems Tools}
Part~2 of the paper uses elementary linear-systems machinery: asymptotics of matrix powers,
spectral radii versus operator norms for non-normal matrices~\cite{trefethen2005spectra}, and
stationary distributions of linear stochastic recursions~\cite{antsaklis2006linear}. We state the
needed facts inline; no background beyond linear algebra is assumed.

\section{Preliminaries}
\label{sec:prelim}

\subsection{Terminology}
A \emph{Database} $\D=\{\Ssub_i\}_{i=1}^{N}$ consists of $N$ distinct subjects (individuals),
where each subject $\Ssub_i=\{I_{i,j}\}_{j=1}^{m_i}$ is simply a collection of $m_i$ facial images
of that person, with each image $I_{i,j}\in\R^{H\times W\times 3}$ given by its pixel
representation. When the subject index is clear from context, we simplify this to
$\Ssub=\{I_j\}_{j=1}^{m}$. Throughout, superscripts mark a subject's \emph{role} (donor $d$,
target $t$, swap $s$, candidate $c$) while subscripts are reserved for numbering (e.g.\ $i$,
$j$). In this setting, a \emph{donor} $\Ssub^d$ is a subject whose face we
have permission to use --- either because their identity is publicly known or because they have
given explicit consent --- while a \emph{target} $\Ssub^t$ is a subject whose identity we aim to
protect. We say that two subjects $\Ssub_1,\Ssub_2$ are the same, i.e.\ $\Ssub_1\equiv\Ssub_2$,
if the images contained in $\Ssub_1$ and $\Ssub_2$ belong to the same person.

\begin{figure}[t]
\centering
\includegraphics[width=\columnwidth]{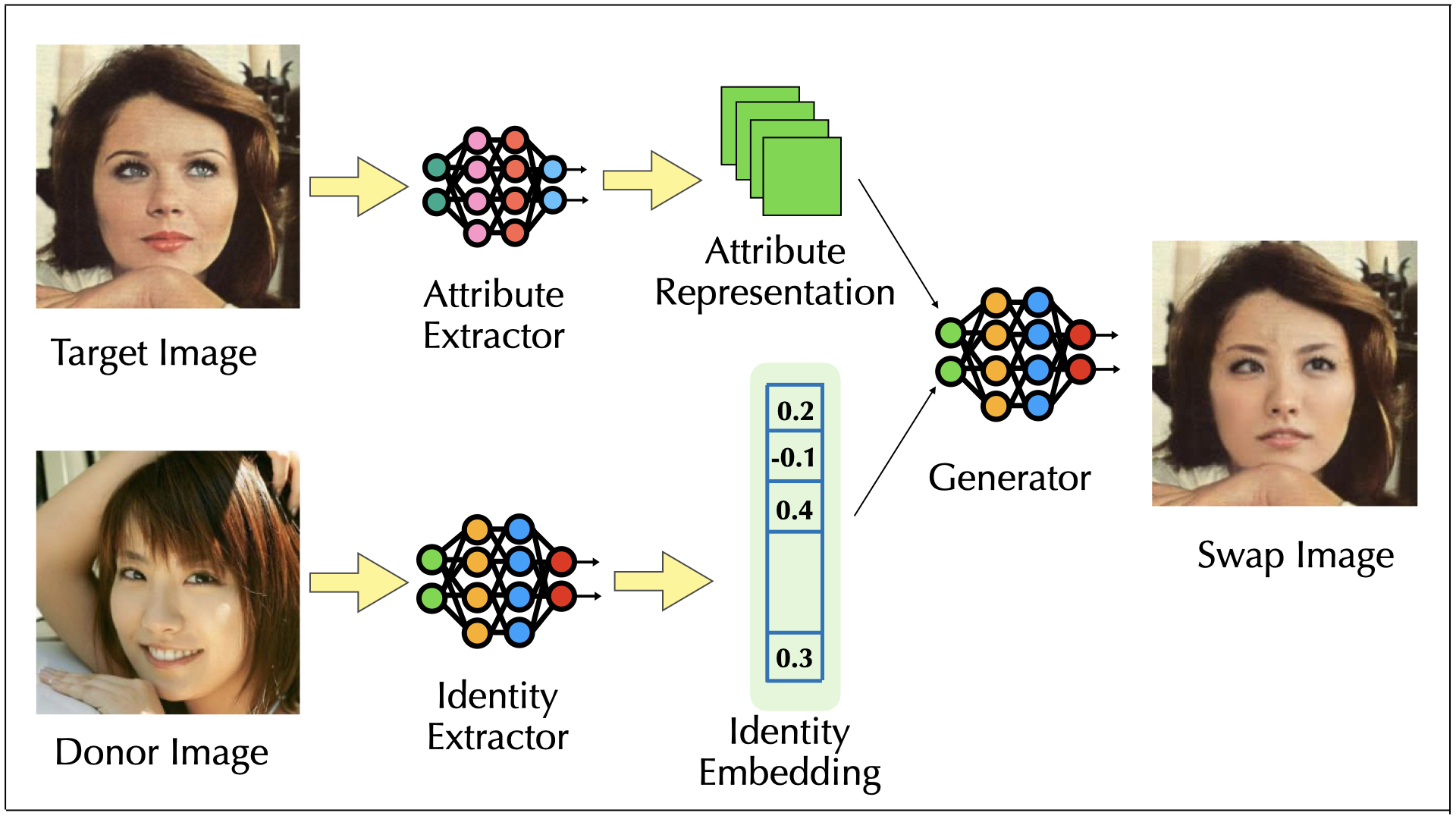}
\caption{General face-swapping workflow.}
\label{fig:swapworkflow}
\end{figure}

\subsection{Identity Extraction}
\label{sec:prelim-extract}
Identity extractor models, based on ArcFace~\cite{deng2019arcface} and
FaceNet~\cite{schroff2015facenet}, are foundational to modern face recognition systems. Their
objective is to map a facial image $I$ into a compact latent representation, or identity
embedding $z$. These identity embeddings capture structural facial characteristics unique to a
person, making them resistant to variations in lighting, pose, and expression. In this embedding
space, images of the same individual cluster closely together, while images of different
individuals remain separated, as in Appendix Figure~\ref{fig:embclusters}. This process utilizes
a trained
neural network backbone, such as ResNet~\cite{he2016deep}, whose raw output features are
subsequently unit-normalized to generate the final identity embedding.

Formally, given an identity extractor $\G:\R^{H\times W\times 3}\to\mathbb{S}^{k-1}$ using a
backbone network $\phi:\R^{H\times W\times 3}\to\R^{k}$, the embedding of an image $I$ is given
by:
\begin{equation}
  z=\G(I):=\frac{\phi(I)}{\norm{\phi(I)}_2}\in\mathbb{S}^{k-1},
\end{equation}
where $\mathbb{S}^{k-1}$ denotes the $k$-dimensional unit hypersphere embedding space. The
similarity $\sigma(z_1,z_2)$ of two embeddings $z_1,z_2\in\mathbb{S}^{k-1}$ is calculated using
cosine similarity as follows:
\begin{equation}
  \sigma(z_1,z_2):=z_1^{\top}z_2\in[-1,1].
\end{equation}

To measure the similarity between an image $I^s$ --- possibly belonging to a synthetically
generated subject --- and a subject $\Ssub=\{I_j\}_{j=1}^{M}$, we compute pairwise similarities
against every image in $\Ssub$,
\begin{equation}
  \sigma_j=\sigma\big(\G(I^s),\G(I_j)\big),\qquad j\in[M].
\end{equation}
These $M$ scores admit two natural summaries, each reflecting a different notion of subject-level
similarity. The \emph{median similarity} $\sigma_{\mathrm{med}}(\Ssub,I^s)=\mathrm{median}_j\{\sigma_j\}$
captures a typical match, robust to the inevitable variation in appearance across images of the
same person. The \emph{max similarity} $\sigma_{\mathrm{max}}(\Ssub,I^s)=\max_j \sigma_j$ instead
asks for the closest image in $\Ssub$ to $I^s$, a best-case measure of identity alignment. 
\subsection{Face-Swapping}
\label{sec:prelim-swap}
Face swapping is the task of transferring the identity of a donor subject $\Ssub^d$ from a source
image $I^d$ onto a target subject $\Ssub^t$ in a target image $I^t$. This process aims to preserve
identity-irrelevant attributes such as pose and expression. Various face-swapping
frameworks~\cite{mirsky2021creation} have been developed using diverse deep learning
architectures, including encoder-decoders~\cite{perov2020deepfacelab}, Generative Adversarial
Networks (GANs)~\cite{goodfellow2014generative,li2020faceshifter}, and diffusion
models~\cite{ho2020denoising,zhao2023diffswap}. The baseline definition and the workflow
illustrated in Figure~\ref{fig:swapworkflow} serve as a general outline; example outputs of the
seven tools studied in this paper are shown in Appendix Figure~\ref{fig:swapexamples}. Individual
models may
diverge from this description depending on their specific architectural designs and
implementations. Furthermore, we do not rigorously define what constitutes an identity-irrelevant
attribute --- such as whether it includes skin tone, moles, blemishes, or lesions. However, we
explicitly assume that background elements, hair, and the neck are entirely preserved from the
target image, while expressions are retained to a reasonable extent. This flexibility exists
because models can be engineered for context-dependent scenarios. For simplicity, we consider face
swapping achieved when a facial recognition identity embedding model confirms the donor's face has
been transferred to the target image.

Formally, a face-swapping model is represented as
\begin{equation}
  \F:\R^{H\times W\times 3}\times\R^{H\times W\times 3}\to\R^{H\times W\times 3},
\end{equation}
which takes a donor image $I^d\in\Ssub^d$ and a target image $I^t\in\Ssub^t$ and produces a swap
image $I^s=\F(I^d,I^t)$.

We further define the action of the face-swapping model within the identity embedding space as:
\begin{equation}
  \FG(z^d,z^t):=\G\big(\F(I^d,I^t)\big),
\label{eq:fg}
\end{equation}
where $z^d=\G(I^d)$ and $z^t=\G(I^t)$ represent the donor and target identity embeddings,
respectively, and $z^s$ denotes the resulting swap embedding. For notational convenience
throughout the remainder of this work, we drop the $z$ prefix and simply write $\FG(d,t)=s$.

\subsection{Identity Inference Attacks}
\label{sec:prelim-iia}
Membership Inference Attacks (MIAs) are a well-established privacy threat in machine learning,
where an adversary seeks to determine whether a particular data sample was used during model
training~\cite{shokri2017membership,yeom2018privacy,salem2019mlleaks,carlini2022membership}.
Inspired by this paradigm, we define the \emph{Identity Inference Attack} (IIA), whose objective
is to determine whether a candidate individual is represented within a privatized database. Given
a candidate subject $\Ssub^c$, a database $\D_p=\{\Ssub_j\}_{j=1}^{N}$ privatized using a
face-swapping model $\F$, an identity inference query can be defined as
\begin{equation}
  Q_{\mathrm{identity}}:(\Ssub^c,\D_p,\F)\to\{0,1\},
\end{equation}
where an output of $1$ indicates that the candidate's identity is represented in the privatized
database, i.e.,
\begin{equation}
  \exists\,\Ssub_j\in\D_p \;\big|\; \Ssub^c\equiv\Ssub_j,
\end{equation}
and an output of $0$ indicates otherwise.

We instantiate this query as a similarity-and-threshold procedure
and refer to the resulting adversarial attack as an Identity Inference Attack (IIA). Given a
candidate subject $\Ssub^{c}$, a privatized database
$\D_{p}=\{\Ssub_i\}_{i=1}^{N}$, and a subject-level similarity $\sigma(\cdot,\cdot)$ built from
the summaries of Section~\ref{sec:prelim-extract},
the adversary first identifies the most similar subject in the database:
\begin{equation}
  \hat{j}=\arg\max_{j\in\{1,\dots,N\}} \sigma\big(\Ssub^{c},\Ssub_j\big).
\end{equation}
The adversary then performs a binary decision based on the similarity score,
\begin{equation}
  \ind{\,\sigma\big(\Ssub^{c},\Ssub_{\hat{j}}\big)\geq\tau\,},
\end{equation}
where $\tau$ is a decision threshold and $\ind{\cdot}$ denotes the indicator function. An output
of $1$ indicates that the adversary infers the candidate identity to be represented in the
privatized database, while an output of $0$ indicates otherwise.

In practice, we instantiate $\sigma(\Ssub^c,\Ssub_j)$ using the subject-level similarity summaries
of Section~\ref{sec:prelim-extract}: each privatized image is embedded once, scored against the
candidate's image gallery, and the per-image scores are aggregated by the median (default) or max.
Sweeping the threshold $\tau$ traces out the attack's ROC curve; the candidate set is split into
true targets (members) and unrelated individuals (non-members) to measure it. We also evaluate a
\emph{closed-set} variant of the same adversary in Appendix~\ref{app:reid}: there, the
adversary holds a gallery of candidate identities known to contain the true target and must
\emph{rank} them, which is the classical person re-identification setting evaluated with CMC
curves.

\section{Threat Model}
\label{sec:threat}

\subsection{Data Owner}
A Data Owner (such as a hospital) possesses a database $\D_{\mathrm{original}}$ of different
target subjects. The Data Owner has accumulated multiple facial images $I^{t}_{i,j}$ of each
target subject $\Ssub^{t}_{i}$ and wishes to release this database after privatization. They use a
public database $\D_{\mathrm{public}}$ of donor subjects $\Ssub^{d}_{i}$ and a face-swapping model
$\F$ to generate a privatized database $\D_{\mathrm{private}}$ in the following manner:
\begin{equation}
  \D_{\mathrm{private}}=\F\big(\D_{\mathrm{public}},\D_{\mathrm{original}}\big),
\end{equation}
where each target image $I^{t}_{i,j}$ is assigned a donor image $I^{d}_{i,j}$ belonging to a
donor subject drawn at random from the public pool, and is replaced by its swap:
\begin{equation}
  \Ssub^{p}_{i}=\big\{\F\big(I^{d}_{i,j},\,I^{t}_{i,j}\big)\big\}_{j=1}^{m_i},
  \qquad
  \D_{\mathrm{private}}=\{\Ssub^{p}_{i}\}_{i=1}^{N}.
\end{equation}
The Data Owner's goal is twofold: \emph{utility} --- released images remain photorealistic faces
preserving the originals' non-identity content (pose, expression, context), the reason
face-swapping was chosen over blurring or inpainting --- and \emph{privacy} --- the release should
reveal nothing about which individuals were present in $\D_{\mathrm{original}}$. The donors'
identities are, by assumption, safe to expose.

\subsection{Adversary}
The adversary observes the released database $\D_{\mathrm{private}}$ and, for each candidate
subject $\Ssub^c$ in a gallery (a handful of ordinary photographs, e.g.\ scraped from social
media), decides whether that person is represented in the release --- the identity inference
query $Q_{\mathrm{identity}}$ of Section~\ref{sec:prelim-iia}. We assume only black-box access to
a strong public face recognizer $\G$ (we use InsightFace \texttt{buffalo\_l}; the adversary need
not know which recognizer, if any, the tool used internally); knowledge that the release was
face-swapped, but not the donor-to-target assignment nor the originals
$\D_{\mathrm{original}}$; and possibly several candidate photographs per identity for
subject-level aggregation. This is a weak --- and therefore realistic --- adversary that trains
nothing and needs only one embedding and a threshold, so any leakage it exploits lower-bounds
what stronger, tool-aware adversaries could extract. A successful attack is a membership
disclosure: learning a person is present in a released clinical archive may itself reveal a
diagnosis.

\subsection{Privacy Goal}
The mechanism is private if the adversary's advantage over random guessing is negligible: the
score distribution a candidate induces against the release should be statistically
indistinguishable whether or not the candidate is a true target. We measure ROC-AUC of
target-vs-non-member discrimination, TPR at low FPR ($\leq 1\%$, $\leq 0.1\%$; average-case
metrics understate harm~\cite{carlini2022membership}), and closed-set rank-$k$ rates. The
non-member score distribution is central: it is both the chance baseline and, as Part~2 shows,
the floor that repeated swapping approaches.

\section{Part 1: Does Face-Swapping Leak Privacy?}
\label{sec:part1}

This section establishes, empirically and across a broad range of tools, the phenomenon that the
rest of the paper explains: a single face-swap suppresses the target's identity but does not
remove it. The test logic is simple. Under an ideal privacy-preserving face-swapping mechanism,
the embedding of a swap image would contain no information about the original target identity,
so the swap's similarity to the true target would be statistically indistinguishable from its
similarity to an unrelated non-member, and any classifier separating the two would achieve an
AUC of approximately $0.5$. We measure how far reality departs from this ideal under a
controlled multi-tool benchmark.

\subsection{Methodology}
\label{sec:part1-method}

\noindent\textbf{Dataset.} We use VGGFace2-HQ~\cite{simswapplusplus}, the published
quality-normalized release of VGGFace2~\cite{cao2018vggface2} in which
all images are pre-aligned and restored to $512\times512$ resolution with
GFPGAN~\cite{wang2021gfpgan}. This removes resolution and alignment artifacts as confounders and
lets every swapping tool operate on clean inputs. The corpus contains 8{,}624 identities with
roughly 1.16M images.

\smallskip
\noindent\textbf{Pair protocol.} From identities with at least 30 images we sample three \emph{disjoint} sets of 1{,}000 identities each: donors, targets, and non-members.
Each experimental unit is a triple (donor, target, non-member): one anchor image of
the donor is swapped onto one anchor image of the target, and the non-member serves as the matched
negative control for that swap. One swap is generated per pair per tool.

\smallskip
\noindent\textbf{Tools.} We evaluate seven publicly available face-swapping systems spanning the
architecture families of Section~\ref{sec:related}: FaceFusion
(\texttt{hyperswap\_1a\_256})~\cite{facefusion}, BlendFace~\cite{shu2023blendface},
CanonSwap~\cite{luo2025canonswap}, DiffSwap~\cite{zhao2023diffswap},
DiffFace~\cite{kim2025diffface}, E4S~\cite{liu2023e4s}, and
FaceShifter~\cite{li2020faceshifter}. Each tool is run with its published checkpoint and default
settings.\footnote{FaceShifter has no official public checkpoint; we trained our own using the
open-source implementation at \url{https://github.com/maum-ai/faceshifter} on
CelebA-HQ~\cite{liu2015celeba}. Its
results should therefore be read as representative of this reimplementation rather than of the
original paper's model.} After removing pairs for which any tool failed (face not detected,
generation error),
\npairs{} complete pairs remain with valid swaps from \emph{all} seven tools; all analyses use
this common subset so that tools are compared on identical inputs. Appendix
Figure~\ref{fig:swapexamples} shows example outputs.

\smallskip
\noindent\textbf{Scoring.} Each swap image is embedded once and scored against three galleries
using the subject-level summaries of Section~\ref{sec:prelim-extract}: the \emph{target gallery}
(all other images of the target identity, excluding the exact image used in the swap), the
\emph{non-member gallery} (images of the assigned non-member identity), and the \emph{donor
gallery} (images of the donor identity). We report the median aggregation in the main text;
max-aggregation results are similar and appear in Table~\ref{tab:mia-metrics}. To control for
recognizer-specific artifacts we repeat all measurements with two independent embeddings:
InsightFace \texttt{buffalo\_l} (ArcFace)~\cite{insightface,deng2019arcface} and
Facenet512~\cite{schroff2015facenet}, the latter as implemented in the DeepFace
library~\cite{serengil2020lightface}.

\smallskip
\noindent\textbf{Attack instantiation.} The membership-style IIA of
Section~\ref{sec:prelim-iia} reduces, under this protocol, to thresholding the subject-level
similarity between a candidate gallery and a released swap: target scores form the positive class
and non-member scores the negative class. Sweeping $\tau$ yields ROC curves, from which we report
AUC, TPR at FPR $\leq 1\%$, and TPR at FPR $\leq 0.1\%$.

\subsection{Donor Identity Transfer: Which Tools Anonymize at All?}
\label{sec:part1-donor}
An anonymizer must first do its job: the released face should carry the \emph{donor's} identity.
Figure~\ref{fig:histgrid} shows, per tool, the distributions of swap similarity to the donor,
target, and non-member galleries; the donor distributions (green) make the comparison immediate.

The field splits cleanly. Five of the seven tools --- FaceFusion, DiffFace, BlendFace,
E4S, and CanonSwap --- produce outputs that are donor-dominated in more than $93\%$ of pairs, with
donor similarities several times larger than target similarities; the remaining few percent
trace largely to difficult inputs (extreme pose, occlusion, low quality)
or occasional swap-model failures. These are the tools a data
owner would plausibly deploy: their outputs would pass a donor-centric or verification-style
audit. The remaining two fail as anonymizers outright: DiffSwap's outputs are closer to the
\emph{target} than to the donor in $86\%$ of pairs, and FaceShifter sits near the boundary
(target similarity $0.198$ vs donor $0.226$) --- under our working
definition of face swapping (Section~\ref{sec:prelim-swap}), these two frequently do not
achieve the swap at all on this data. We therefore treat the five donor-retaining tools as the
relevant population for anonymization and restrict the mechanistic analysis of Part~2 to
tools from this population; DiffSwap and FaceShifter remain in the leakage tables as
cautionary upper bounds, their high leakage unsurprising --- the target was never fully
removed.

Note also the non-member column: unrelated identities score essentially zero (means
$0.004$--$0.009$) against swaps under \texttt{buffalo\_l}, so whatever target similarity remains
is target-specific signal, not generic face-likeness.

\subsection{Target Leakage: Score Distributions}
\label{sec:part1-hist}
\begin{figure*}[!t]
\centering
\includegraphics[width=0.97\textwidth]{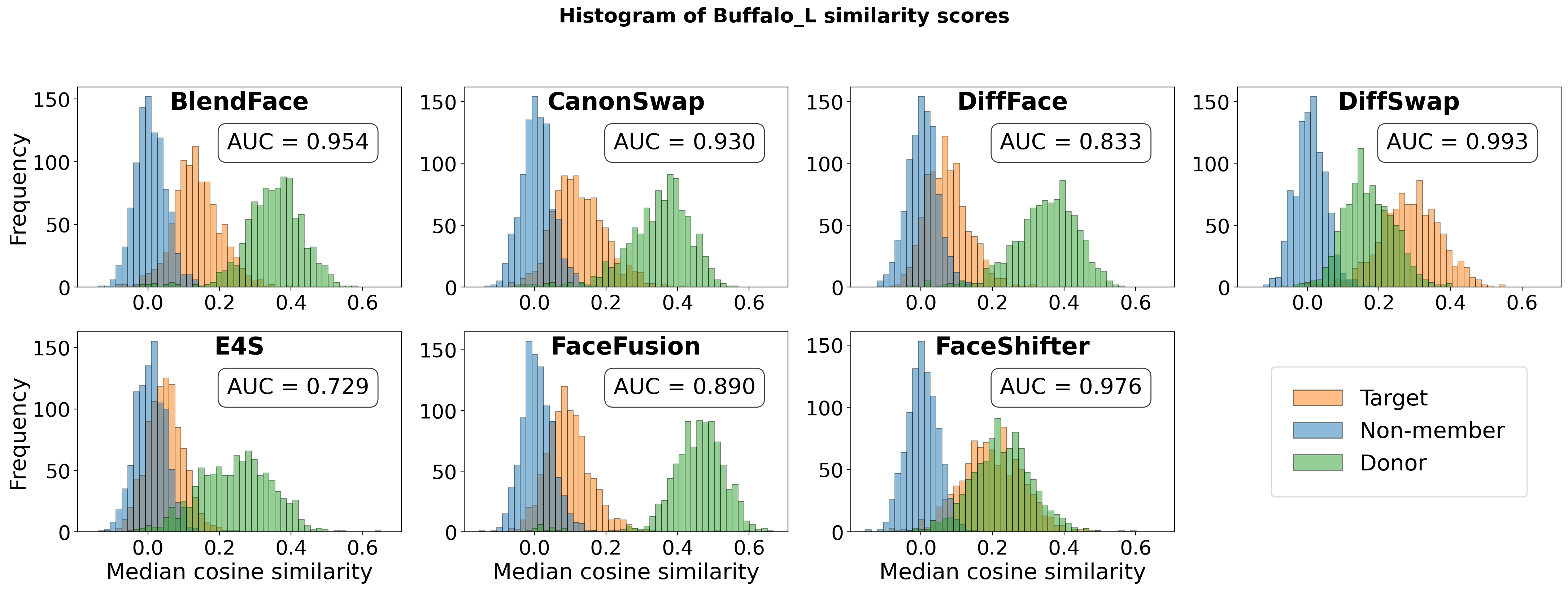}
\caption{Single-swap similarity distributions (VGGFace2, \texttt{buffalo\_l}, median gallery
aggregation, \npairs{} pairs per tool). Each panel is one tool; histograms compare swap
similarity to the target (orange), an unrelated non-member (blue), and the donor (green). The
reported AUC is target-vs-non-member discrimination. For every tool the target distribution is
shifted away from the non-member baseline: a single swap suppresses, but does not erase, the
target identity.}
\label{fig:histgrid}
\end{figure*}

Figure~\ref{fig:histgrid} overlays, for each tool, the distributions of target, non-member, and
donor scores across the \npairs{} pairs. The non-member distribution (blue) is tightly
concentrated near zero, as expected for unrelated identities. Under ideal anonymization the
target distribution (orange) would coincide with it. Instead, for every tool, the target
distribution is visibly shifted to the right, with per-tool target-vs-non-member AUCs ranging
from $0.729$ (E4S) to $0.993$ (DiffSwap). The donor distribution (green) confirms
the split of Section~\ref{sec:part1-donor}: well-separated and high for the five anonymizing
tools, collapsed onto the
target distribution for FaceShifter, and inverted for DiffSwap.

Two features deserve emphasis. First, suppression is real and large: for
FaceFusion the mean target score drops from $0.616$ (a genuine target image against
its own gallery, cf.\ Section~\ref{sec:verify}) to $0.098$ --- an $84\%$ reduction that a
verification-style evaluation would call success. Second, the suppressed scores remain
\emph{systematically} above the non-member baseline of $\approx 0.008$, by an order
of magnitude in the mean. Privacy is a distributional property, and distributionally the target
is still there.

\subsection{Membership Inference: ROC and TPR at Low FPR}
\label{sec:part1-mia}
\begin{figure}[t]
\centering
\includegraphics[width=\columnwidth]{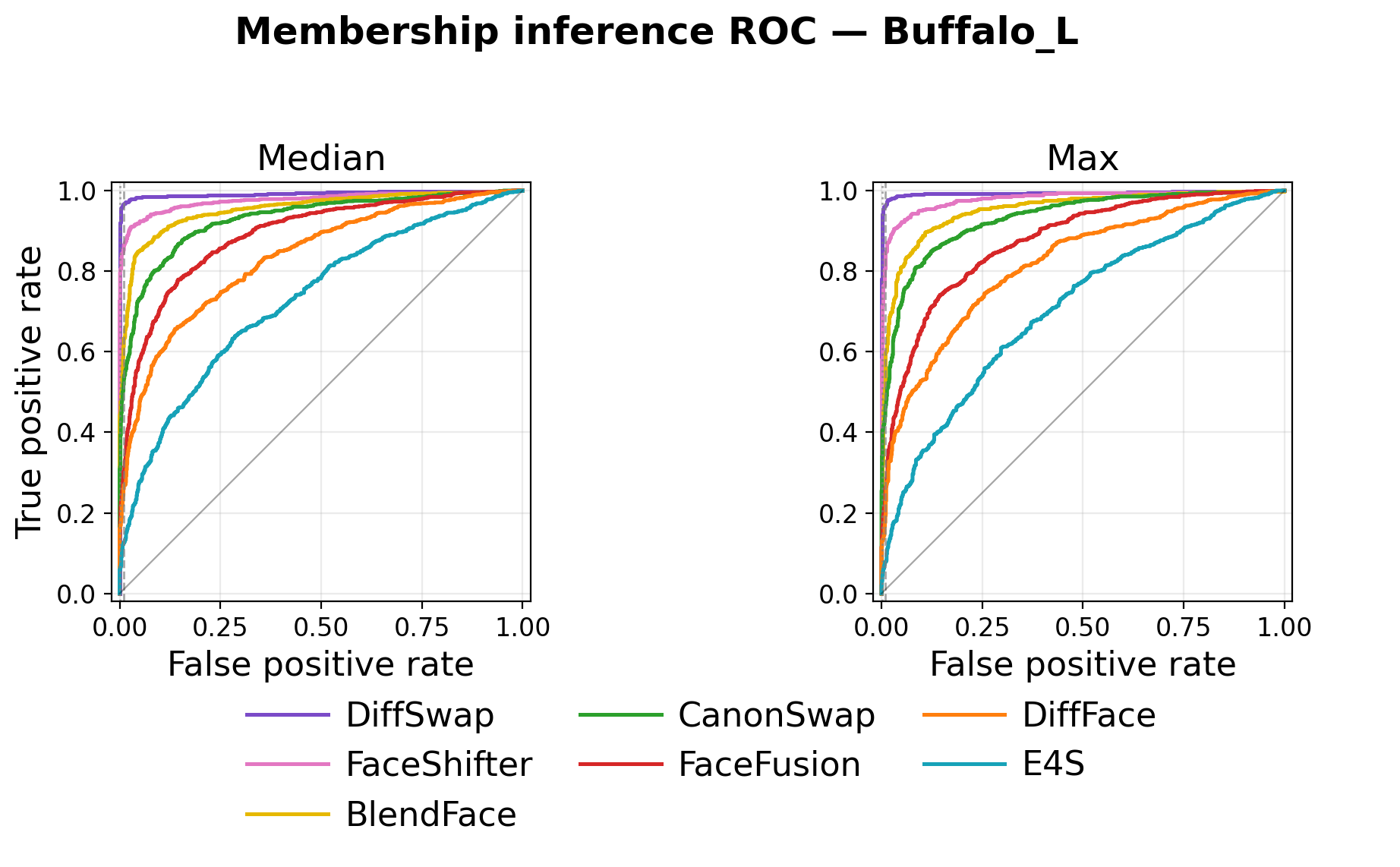}
\caption{Membership-style identity inference ROC curves (\texttt{buffalo\_l}; left: median
aggregation, right: max). The adversary thresholds a single cosine score and remains far above
chance for every tool after one swap.}
\label{fig:roc}
\end{figure}

Figure~\ref{fig:roc} shows the ROC curves of the thresholding adversary, and
Table~\ref{tab:mia-metrics} reports the operating points that matter
for privacy: TPR at FPR
$\leq 1\%$ and $\leq 0.1\%$, for both recognizers and both aggregation rules.

\begin{table*}[t]
  \centering
  \caption{Membership inference attack metrics on VGGFace2 (seed 42, \npairs{}--948 pairs per
    tool). Positive class: target (member) similarity; negative: non-member. Higher similarity
    score predicts membership. TPR values are percentages. Best results (lowest leakage) are in
    \textbf{bold}; worst (highest leakage) are \underline{underlined}. The horizontal rule
    separates the five donor-retaining tools from the two that fail to transfer donor identity
    (Section~\ref{sec:part1-donor}).}
  \label{tab:mia-metrics}
  \small
  \setlength{\tabcolsep}{5pt}
  \begin{tabular}{lcccccccccccc}
    \toprule
    & \multicolumn{4}{c}{TPR @ FPR $\leq 1\%$}
    & \multicolumn{4}{c}{TPR @ FPR $\leq 0.1\%$}
    & \multicolumn{4}{c}{Max balanced acc.} \\
    \cmidrule(lr){2-5} \cmidrule(lr){6-9} \cmidrule(lr){10-13}
    & \multicolumn{2}{c}{Buffalo\_L} & \multicolumn{2}{c}{Facenet512}
    & \multicolumn{2}{c}{Buffalo\_L} & \multicolumn{2}{c}{Facenet512}
    & \multicolumn{2}{c}{Buffalo\_L} & \multicolumn{2}{c}{Facenet512} \\
    \cmidrule(lr){2-3} \cmidrule(lr){4-5} \cmidrule(lr){6-7} \cmidrule(lr){8-9} \cmidrule(lr){10-11} \cmidrule(lr){12-13}
    Tool & Med. & Max & Med. & Max & Med. & Max & Med. & Max & Med. & Max & Med. & Max \\
    \midrule
    BlendFace   & 61.3 & 59.8 & 26.6 & 24.1 & 44.1 & 16.5 & 14.2 &  4.6 & 90.1 & 89.4 & 81.8 & 78.9 \\
    CanonSwap   & 51.9 & 46.3 & 27.0 & 21.8 & 30.8 & 25.5 & 10.8 &  9.9 & 86.1 & 86.3 & 77.2 & 75.2 \\
    DiffFace    & 26.0 & 19.4 & 15.7 & 13.2 & 16.1 & 11.4 &  6.0 &  \textbf{1.1} & 76.0 & 74.4 & 71.5 & 69.9 \\
    E4S         & \textbf{12.4} &  \textbf{7.8} &  \textbf{6.9} &  \textbf{5.4} &  \textbf{5.8} &  \textbf{2.1} &  \textbf{0.4} &  1.4 & \textbf{67.6} & \textbf{65.7} & \textbf{62.7} & \textbf{60.9} \\
    FaceFusion  & 30.2 & 26.3 & 15.5 & 13.4 & 14.6 & 11.6 &  3.6 &  4.5 & 81.7 & 79.8 & 72.9 & 72.1 \\
    \midrule
    DiffSwap    & \underline{96.7} & \underline{96.4} & \underline{86.8} & \underline{79.6} & \underline{90.8} & \underline{78.2} & \underline{61.7} & \underline{59.0} & \underline{97.9} & \underline{98.1} & \underline{94.8} & \underline{94.4} \\
    FaceShifter & 85.2 & 84.7 & 44.2 & 27.6 & 72.6 & 58.2 & 11.9 &  9.3 & 94.1 & 93.8 & 81.9 & 80.3 \\
    \bottomrule
  \end{tabular}
\end{table*}

The low-FPR columns are the ones to read. At FPR $\leq 1\%$ --- the adversary is essentially never
wrong about a non-member --- BlendFace gives up $61\%$ of its protected targets and CanonSwap
$52\%$. Even FaceFusion, the strongest donor-transferer, yields $30\%$
of targets at $1\%$ FPR and $15\%$ at $0.1\%$ FPR: roughly one in seven protected individuals
confidently identified with a false-alarm rate of one in a thousand. E4S is the most private
tool at every operating point ($12.4\%$ TPR at $1\%$ FPR), though it also writes the donor in
most weakly ($0.246$ mean donor similarity), suggesting its
advantage comes partly from synthesizing a face far from \emph{both} input identities.
The weaker Facenet512 recognizer sees less leakage at every operating point but preserves the
tool ordering: leakage is a property of the swaps, not of one
embedding, and a better recognizer extracts more of it. Aggregation matters less, though
median aggregation is the stronger
attack at $0.1\%$ FPR ($44.1\%$ vs $16.5\%$ for BlendFace). A \emph{closed-set} variant of
this adversary, which ranks all 948 candidate identities instead of thresholding, is starker
still --- rank-1 identification reaches $34\%$ of swaps for BlendFace versus $0.1\%$ chance; full
protocol, CMC curves, and rank-$k$ results are given in Appendix~\ref{app:reid}.

\subsection{Robustness Checks}
\label{sec:part1-robust}
\noindent\textbf{Harder non-members.} One may object that random non-members are too easy a
negative class: a demographically similar non-member would score higher and shrink the gap. We
re-ran the benchmark replacing each pair's non-member with an adversarially chosen one: the
candidate, among all VGGFace2 identities sharing the target's annotated attributes
(MAAD-Face~\cite{terhorst2021maad}), with the highest embedding similarity to the target. As
expected, non-member scores rise substantially (e.g.\ from $0.008$ to $0.041$ mean for BlendFace,
$0.029$ for FaceFusion), while target scores are unchanged; the target-vs-non-member gap narrows
but persists for every tool, and the tool ordering is identical. Demographic matching explains
part of the baseline, not the leakage.

\smallskip
\noindent\textbf{Second recognizer.} All effects replicate under Facenet512
(Table~\ref{tab:mia-metrics}), an embedding trained with a different loss, architecture, and
dataset from any recognizer used inside the swapping tools. Leakage is therefore not an artifact
of shared training between the attack recognizer and the tools' internal identity encoders.

\subsection{Summary of Part 1}
Single-pass face-swapping, as implemented by every modern tool we tested, leaks target identity:
suppressed match scores remain an order of magnitude above the non-member baseline, supporting
membership inference at AUC $0.73$--$0.95$ for the five tools that anonymize at all, with TPR up
to $61\%$ at $1\%$ FPR, and closed-set re-identification up to $34\%$ rank-1 out of 948
candidates. The effect is robust to the recognizer, the aggregation rule, and adversarial
non-member selection. Leakage varies by an order of magnitude across tools, which raises the
questions Part~2 answers: where does the surviving target signal live, why does its magnitude
differ across tools, and what happens when the defense is iterated?

\section{Part 2: Face-Swappers as Affine Stochastic Operators}
\label{sec:model}
Part~1 measured leakage; this part models it. The object we model is not the image-space
generator $\F$ but its shadow $\FG$ on the adversary's identity-embedding space
(Equation~\ref{eq:fg}). Privacy lives or dies in this space --- every attack in Part~1 was a
function of embeddings alone --- so a faithful model of $\FG$ is exactly what is needed to
explain and predict leakage.

\subsection{Model Definition}
\label{sec:model-def}
We posit that, to first order, the swap embedding is an affine combination of the donor and
target embeddings plus a stochastic innovation:
\begin{equation}
  \bm{s}=\Amat\bm{d}+\Bmat\bm{t}+\cvec+\evec,
\label{eq:swap}
\end{equation}
where $\bm{d},\bm{t},\bm{s}\in\R^{512}$ are donor, target, and swap embeddings,
$\Amat,\Bmat\in\R^{512\times512}$ are the \emph{donor-write} and \emph{target-transfer}
operators, $\cvec\in\R^{512}$ is a bias, and $\evec$ is a zero-mean residual capturing
everything the affine part misses: pose and expression effects, generator idiosyncrasies, and
blending artifacts. The two operators have direct privacy interpretations. $\Amat$ measures how
strongly the swapper writes the donor into the output --- its job. $\Bmat$ measures how much of the
target passes through --- the leakage channel. If face-swapping were a perfect anonymizer in
embedding space, $\Bmat$ would be the zero matrix and Part~1 would have found nothing.

Three remarks set expectations. First, \eqref{eq:swap} is \emph{not} a claim that the
image-space generator is linear; it is a first-order regression model of the recognizer's view
of the generator, justified empirically below. Second, the model is intentionally the simplest
one that can express partial identity transfer; if it fits, the payoff is large, because affine
maps compose, and composition (repeated swapping) is exactly the privacy mechanism we want to
analyze. Third, modern swapper architectures make linearity in this particular space less
surprising than it may seem: the donor is handed to the generator not as pixels but as an
ArcFace-style identity embedding, which conditions synthesis through linear modulation layers
(learned per-channel scales and shifts of generator features).

\subsection{Fitting Protocol}
\label{sec:model-fit}
\noindent\textbf{Corpora.} Fitting a $512\times512$ operator pair requires far more than the
\npairs{} swaps per tool. We therefore generated dedicated large-$N$ triplet corpora
$(\bm{d}_i,\bm{t}_i,\bm{s}_i)$ on VGGFace2: \ffTrain{} training and \ffTest{} held-out test
triplets for FaceFusion, \bfTrain{}/\bfTest{} for BlendFace, and \csTrain{}/\csTest{} for
CanonSwap, all drawn from the same donor/target pair distribution so that operators are
comparable across tools. These three tools are drawn from the five donor-retaining tools
identified in Part~1; they were selected because they span the leakage range of that group
(Figure~\ref{fig:histgrid}) while remaining computationally feasible to run at the
tens-of-thousands-of-swaps scale that operator identification demands (Appendix~\ref{sec:model-corpus}).

\smallskip
\noindent\textbf{Identity-disjoint splits.} Train and test sets share no identities: a triplet
is kept only if both its donor and target identities fall entirely in the train or entirely in
the test partition. Reported fit quality therefore measures generalization to unseen people, not
memorization of identities.

\smallskip
\noindent\textbf{Estimator.} We fit $(\Amat,\Bmat,\cvec)$ by ridge regression, with the
regularization strength chosen by 5-fold cross-validation over identity-disjoint folds of the
training set. We fit in the \emph{raw} (un-normalized) ArcFace feature space
$\phi(I)\in\R^{512}$ rather than on the unit hyper-sphere: the affine recursion that powers the
cascade analysis of Section~\ref{sec:cascade} is exact in a linear space, whereas the sphere is
not closed under affine maps. The bridge back to Part~1's cosine similarities is empirical and
turns out to be benign: raw swap-embedding norms are essentially constant (about $23.2$,
stable across cascade depth; Section~\ref{sec:verify}), so directions --- and hence cosines ---
carry all the identity information.

\smallskip
\noindent\textbf{Metrics.} On held-out triplets we report the mean cosine between predicted and
true swap embeddings, $\cos(\hat{\bm{s}},\bm{s})$, and the coefficient of determination $R^2$
(fraction of swap-embedding variance explained).

\smallskip
\noindent\textbf{Adequacy.} A linear model of a deep generative pipeline should be treated with
suspicion, so before using \eqref{eq:swap} we subjected it to five checks, reported in full in
Appendix~\ref{sec:model-adequacy}. In brief: (i) the affine model beats every simpler predictor
by a wide margin (held-out cosine \fitcos{} vs.\ $0.654$ for copying the donor embedding and
$0.442$ for a $k$NN regressor); (ii) explicit nonlinear feature corrections improve held-out
$R^2$ by only ${\sim}0.001$; (iii) the residual, though large ($\stochfrac$ of swap variance for
FaceFusion), has the scale of natural same-person embedding variation; (iv) it is nearly
uncorrelated with the target identity (mean absolute correlations $\leq0.023$), so it does not
hide a deterministic target channel; and (v) formal tests show residual exogeneity and
approximate unbiasedness hold, while isotropy and Gaussianity fail --- failures that matter only
for fine-grained distributional claims, not for the first-moment decay rates and floors that the
cascade analysis of Section~\ref{sec:cascade} relies on.

\subsection{Fitted Operators Across Tools}
\label{sec:model-ops}
Table~\ref{tab:crossop} reports the fits for all three tools. Two patterns deserve note. First,
fit quality tracks Part~1 architecture intuition: FaceFusion, whose pipeline most directly
conditions on an ArcFace donor embedding, is the most linear ($R^2=\fitRtwo$); BlendFace and
CanonSwap are progressively less so ($\bfRtwo$, $\csRtwo$), with the majority of their
swap-embedding variance left to the residual. The affine model should accordingly be read as a
strong first-order account for FaceFusion and a coarser one for the other two --- a caveat we
carry into the predictions. Second, despite the differing fit quality, all three target-transfer
operators share the structural features that drive the cascade analysis, described next.

\begin{table*}[t]
\centering
\caption{Fitted affine-stochastic operators for three donor-retaining tools (raw ArcFace space,
ridge with identity-disjoint CV, common donor/target pair distribution). $b_u$ is the mean
Rayleigh gain of $\Bmat$ along real target directions; $\rho(\Bmat)$ the spectral radius;
$\sigma_{\max}(\Bmat)$ the largest singular value; ``resid.\ frac.''\ the unexplained variance
fraction. The last column anticipates Section~\ref{sec:verify}: the measured late-pass cascade
decay ratio.}
\label{tab:crossop}
\begin{tabular}{lcccccccc}
\toprule
Tool & Train / Test & Fit cos & $R^2$ & resid.\ frac. & $b_u$ & $\rho(\Bmat)$ & $\sigma_{\max}(\Bmat)$ & Late cascade ratio \\
\midrule
FaceFusion & \ffTrain{} / \ffTest{} & \fitcos & \fitRtwo & \stochfrac & \buGain & \rhoB & \sigmaxB & \mratiolate \\
BlendFace  & \bfTrain{} / \bfTest{} & \bfFitCos & \bfRtwo & 0.600 & \bfBu & \bfRhoB & \bfSigma & $\approx$0.88 \\
CanonSwap  & \csTrain{} / \csTest{} & \csFitCos & \csRtwo & 0.662 & \csBu & \csRhoB & \csSigma & $\approx$0.79 \\
\bottomrule
\end{tabular}
\end{table*}

\begin{figure}[t]
\centering
\includegraphics[width=\columnwidth]{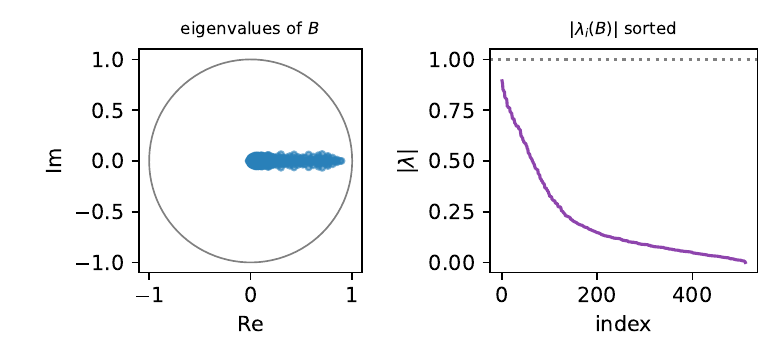}
\caption{Spectrum of the fitted FaceFusion target-transfer operator $\Bmat$. Left: eigenvalues
in the complex plane against the unit circle; all lie strictly inside, so the target carrier
must fade under repeated swapping. Right: sorted eigenvalue moduli; the dominant modulus
$\rho(\Bmat)=\rhoB$ sets the asymptotic decay rate.}
\label{fig:spectrum}
\end{figure}

For FaceFusion, the donor-write operator is strong and expansive along many directions
($\rho(\Amat)=\rhoAval$, with dozens of eigenvalues near or above $1$): the swapper amplifies
donor identity, as it should. The target-transfer operator $\Bmat$ is, by contrast, a
\emph{contraction in spectrum but not in norm}: every eigenvalue lies strictly inside the unit
circle ($\rho(\Bmat)=\rhoB$; Figure~\ref{fig:spectrum}), yet its largest singular value is
$\sigma_{\max}(\Bmat)=\sigmaxB$. This large gap means $\Bmat$ is highly
non-normal~\cite{trefethen2005spectra}: a single application can amplify particular directions
more than tenfold even though repeated application must eventually contract everything. The mean
Rayleigh gain along \emph{real target directions},
$b_u=\mathbb{E}\big[\bm{u}^{\top}\Bmat\bm{u}\big]\approx\buGain$ (where $\bm{u}$ ranges over
unit-normalized test-set target embeddings), is small: in the direction where the protected
identity actually enters, one swap passes through only about $18\%$ of the signal. The tension
between these three numbers --- $b_u\approx0.18$, $\rho(\Bmat)\approx0.89$,
$\sigma_{\max}(\Bmat)\approx10$ --- is not a nuisance; it is the explanation of cascade behavior,
as Section~\ref{sec:cascade} shows.

\section{Cascade Dynamics and Testable Predictions}
\label{sec:cascade}
The natural defense against the leakage of Part~1 is \emph{dilution}: if one swap attenuates the
target, swap again, re-swapping the previous output with a fresh
donor at every pass. The affine model turns this countermeasure into a linear stochastic
dynamical system whose behavior can be predicted before running a single extra swap. This
section derives those predictions; the next one tests them.

\subsection{The Cascade Recursion and the Target Carrier}
\label{sec:cascade-recursion}
Let $\bm{t}_0$ be the embedding of the original protected target and set $\bm{s}_0=\bm{t}_0$. At
pass $k$ the data owner draws a fresh donor $\bm{d}_k$ and swaps it onto the previous output, so
applying \eqref{eq:swap} with the previous output in the target slot gives
\begin{equation}
  \bm{s}_k=\Amat\bm{d}_k+\Bmat\bm{s}_{k-1}+\cvec+\evec_k .
\label{eq:recursion}
\end{equation}
Unrolling the recursion separates the fate of the protected identity from everything else:
\begin{equation}
  \bm{s}_k=\underbrace{\Bmat^{k}\bm{t}_0}_{\text{target carrier}}
  +\sum_{j=0}^{k-1}\Bmat^{j}\big(\Amat\bm{d}_{k-j}+\cvec+\evec_{k-j}\big).
\label{eq:unroll}
\end{equation}
The protected identity enters the cascade exactly once, at pass one, and is thereafter
propagated only by repeated application of the target-transfer operator: the \emph{carrier}
$\Bmat^k\bm{t}_0$. The fresh donors and residuals accumulate in the second term, but --- by the
exogeneity and uncorrelatedness checks of Appendix~\ref{sec:model-adequacy} --- they carry no
information about $\bm{t}_0$. Cascade privacy therefore reduces to a single question of linear
algebra: \emph{what do powers of $\Bmat$ do to the target direction?}

\subsection{Spectral Analysis: Two Rates and a Floor}
\label{sec:cascade-spectral}
\noindent\textbf{Asymptotic rate.} For any matrix, $\norm{\Bmat^k\bm{t}_0}^{1/k}\to\rho(\Bmat)$
as $k\to\infty$: the long-run decay (or growth) of the carrier is governed by the spectral
radius, i.e.\ by eigenvalues. If $\rho(\Bmat)<1$ the carrier fades exponentially at asymptotic
rate $\rho(\Bmat)$ per pass; if $\rho(\Bmat)\geq1$, dilution would fail to erase the target.
All three fitted operators satisfy $\rho(\Bmat)<1$ (Table~\ref{tab:crossop}), so the model
predicts dilution works --- eventually.

\smallskip
\noindent\textbf{Transient rate.} The asymptotic rate is not the first-pass rate. The
\emph{one-step} attenuation of the carrier along a real target direction $\bm{u}$ is the
Rayleigh gain $\bm{u}^{\top}\Bmat\bm{u}\approx b_u\approx0.18$ --- five times smaller than
$\rho(\Bmat)\approx0.89$. The reason is the extreme non-normality of $\Bmat$
($\sigma_{\max}\!\approx\!10\gg\rho$): real target embeddings are far from the dominant
eigenvectors of $\Bmat$, so the first application crushes them; but whatever component survives
is progressively rotated into the slowest-decaying eigenspace, after which decay proceeds at the
spectral rate. The model thus predicts a characteristic \emph{two-rate} signature: a dramatic
first-pass drop (rate $\approx b_u$), then a transition over a few passes, then a slow geometric
tail (rate $\to\rho(\Bmat)$). Because the tail rate is $0.89$
rather than $0.18$, later passes are nearly an order of magnitude less
effective than the first --- a quantitative prediction with direct operational consequences.

\smallskip
\noindent\textbf{Stationary floor.} Because $\rho(\Bmat)<1$, the recursion \eqref{eq:recursion}
is stable and forgets its initial condition: as $k$ grows, $\bm{s}_k$ converges in distribution
to a stationary law with mean and covariance
\begin{align}
  \bm{\mu}_\infty &=(I-\Bmat)^{-1}\big(\Amat\,\mathbb{E}[\bm{d}]+\cvec\big),\\
  \Sigma_\infty &=\Bmat\Sigma_\infty\Bmat^{\top}+\Sigma_w,
\end{align}
where $\Sigma_w=\Amat\,\mathrm{Cov}(\bm{d})\,\Amat^{\top}+\Sigma_e$~\cite{antsaklis2006linear}.
The crucial property is that this distribution \emph{does not depend on} $\bm{t}_0$: a deeply
cascaded swap is statistically just ``some face'', whose similarity to the original
target can neither fall below nor settle above that of an unrelated face.
The leakage plateau is the non-member baseline of Part~1, not zero. Solving the Lyapunov
equation with the fitted FaceFusion operator predicts a plateau of $0.0055$ in median gallery
cosine, against a measured non-member baseline of $\mfloor$.

\subsection{Predictions}
\label{sec:cascade-predictions}
We can now state the model's predictions as falsifiable claims, fixed before examining the
verification data. Define the \emph{excess leakage} at pass $k$ as the median target-gallery
cosine minus the non-member floor, and the \emph{excess ratio} as its ratio between consecutive
passes.

\begin{prediction}[Monotone two-rate fade]
\label{pred:tworate}
Excess leakage decreases monotonically with cascade depth, but not at one rate: the first excess
ratio is small (near $b_u\approx0.18$ for FaceFusion) and subsequent ratios climb toward, and
then track, $\rho(\Bmat)\approx0.89$. The asymptotic ratio matches the spectral radius --- not the
singular value $\sigma_{\max}(\Bmat)\approx10$, which would predict amplification.
\end{prediction}

\begin{prediction}[Floor equals the non-member baseline]
\label{pred:floor}
Leakage plateaus at the non-member similarity baseline rather than zero, and the
membership-inference attack of Part~1 decays toward chance (AUC $0.5$) as depth grows.
\end{prediction}

\begin{prediction}[First pass dominates]
\label{pred:firstpass}
In absolute terms, the first swap removes the large majority of target similarity; all later
passes combined remove far less.
\end{prediction}

\begin{prediction}[Persistent directions]
\label{pred:persistent}
Targets whose embeddings align with the dominant eigenspace of $\Bmat$ decay more slowly across
passes than targets that do not.
\end{prediction}

\begin{prediction}[Cross-tool spectral ordering]
\label{pred:crosstool}
Among tools, fitted spectral radii order the cascade tails: BlendFace
($\rho=\bfRhoB$) should have the slowest late decay, FaceFusion ($\rho=\rhoB$) intermediate,
CanonSwap ($\rho=\csRhoB$) the fastest.
\end{prediction}

\begin{figure*}[!t]
\centering
\includegraphics[width=0.9\textwidth]{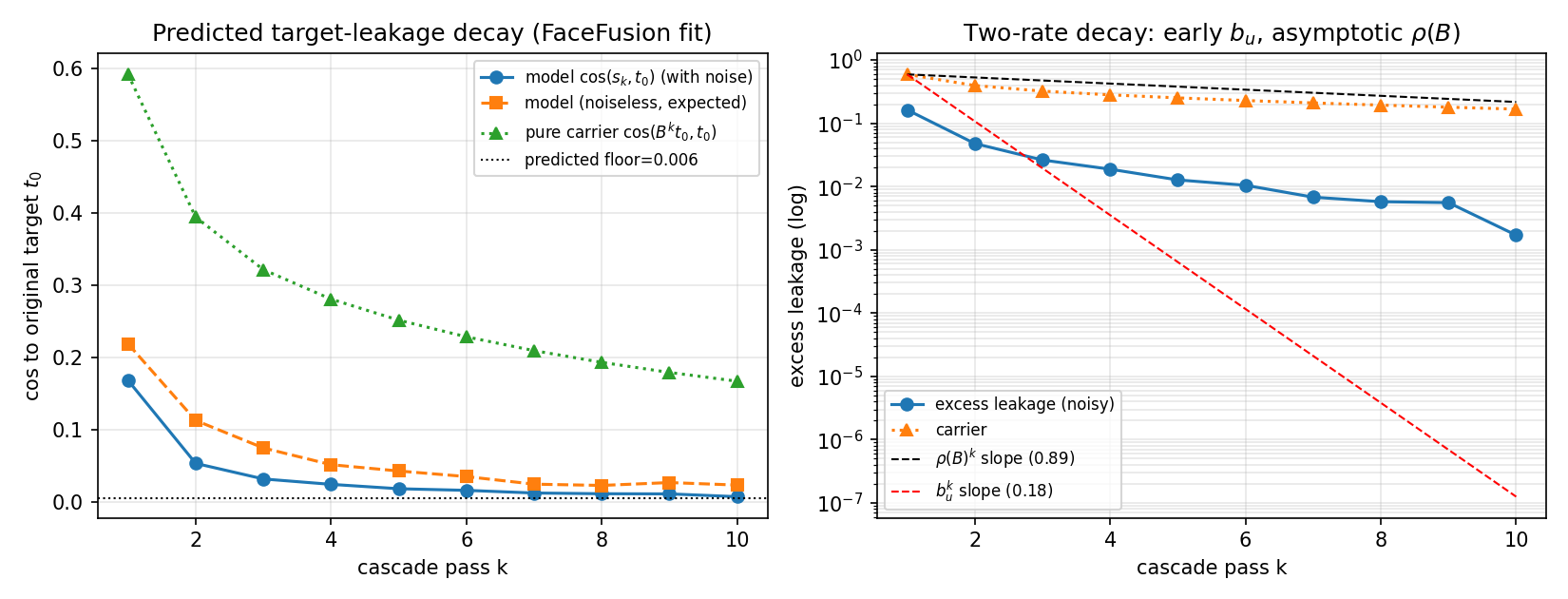}
\caption{Model predictions before verification (FaceFusion fit). Left: simulated target leakage
versus cascade pass for the noisy model, its noiseless expectation, and the pure carrier
$\cos(\Bmat^k\bm{t}_0,\bm{t}_0)$, with the predicted stationary floor. Right: the same
information as excess-leakage slopes on a log scale --- the early decay follows the $b_u$ slope
and bends onto the $\rho(\Bmat)$ slope, the two-rate signature of
Prediction~\ref{pred:tworate}.}
\label{fig:predicted}
\end{figure*}

Beyond these qualitative claims, a Monte Carlo rollout of \eqref{eq:recursion} with the fitted
operator, donor statistics, and residual covariance yields \emph{quantitative} per-pass leakage
estimates (Figure~\ref{fig:predicted}); we will see it gets the rates right and the levels
wrong, an instructive failure dissected below.

\section{Verification on Dilution Data}
\label{sec:verify}

\subsection{A Five-Pass FaceFusion Cascade}
\label{sec:verify-protocol}
To test the predictions at depth we generated a dedicated cascade corpus, designed so that no
protocol artifact could mimic the predicted dynamics. Starting from 500 target identities
(disjoint from all operator-fitting identities), we ran $K=\Kpass$ passes of FaceFusion with a
\emph{fresh donor identity at every pass}: $2{,}500$ distinct donor identities, each
contributing one image, with strict no-reuse of any image anywhere in the experiment.
Each target retains a held-out gallery of up to 30 other images for scoring, and each chain is
paired with a non-member identity as in Part~1. After excluding chains with undetectable faces
mid-cascade, \Nchains{} complete chains remain (Appendix Figure~\ref{fig:dilutionsamples} shows
example chains; visual quality is largely preserved across passes). At every pass we record the
median
target-gallery cosine, the non-member score, and the raw embedding
norm. Pass $0$ denotes the original target image itself, whose gallery score $\selfsim$
calibrates the scale: this is what ``no privacy'' looks like.

\subsection{Results: Decay, Rates, Floor}
\label{sec:verify-results}
\begin{figure}[t]
\centering
\includegraphics[width=\columnwidth]{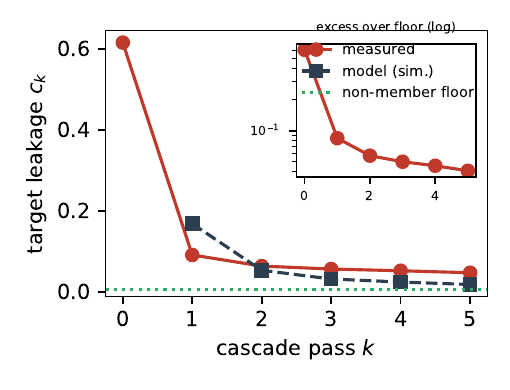}
\caption{Measured target leakage across the \Kpass-pass FaceFusion cascade (median over
\Nchains{} chains), the model rollout, and the non-member floor. Inset: excess over the floor on
a log scale. Leakage fades toward the floor, but the measured tail sits above the simulated
one.}
\label{fig:decay}
\end{figure}

Table~\ref{tab:cascade} and Figure~\ref{fig:decay} present the measured series alongside the
Monte Carlo prediction.

\begin{table}[t]
\centering
\caption{Measured five-pass FaceFusion cascade (\Nchains{} chains, median over chains) versus
the Monte Carlo model rollout. ``Excess ratio'' is the pass-to-pass ratio of leakage above the
non-member floor ($\mfloor$); the model's asymptotic prediction for this ratio is
$\rho(\Bmat)=\rhoB$.}

\label{tab:cascade}
\resizebox{\columnwidth}{!}{%
\begin{tabular}{lccccc}
\toprule
Pass & Measured & Predicted & Non-mem. & Excess ratio & MIA AUC \\
\midrule
0 & 0.616 & ---   & ---   & ---   & --- \\
1 & 0.091 & 0.168 & 0.006 & 0.139 & 0.874 \\
2 & 0.064 & 0.053 & 0.003 & 0.677 & 0.804 \\
3 & 0.056 & 0.032 & 0.008 & 0.873 & 0.761 \\
4 & 0.052 & 0.024 & 0.004 & 0.915 & 0.762 \\
5 & 0.047 & 0.018 & 0.011 & \textbf{0.893} & 0.709 \\
\bottomrule
\end{tabular}}
\end{table}

\begin{figure}[t]
\centering
\includegraphics[width=0.9\columnwidth]{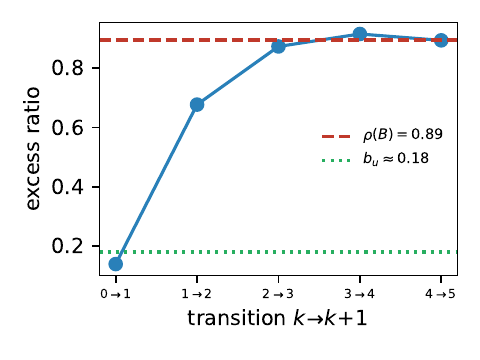}
\caption{Per-pass excess-leakage ratios in the measured cascade, climbing from the
$b_u\approx0.18$ regime toward the spectral rate. The late ratio ($\mratiolate$) coincides with
the independently fitted $\rho(\Bmat)=\rhoB$.}
\label{fig:ratio}
\end{figure}

\noindent\textbf{Prediction~\ref{pred:tworate} (two-rate fade): confirmed in rate, with one
caveat.} The measured leakage decreases monotonically at every pass, and the excess ratios climb
exactly as predicted: $0.139\to0.677\to0.873\to0.915\to0.893$
(Figure~\ref{fig:ratio}). The first ratio ($\mratioearly$) is in the predicted $b_u$ regime, and
the late ratio is $\mratiolate$ --- matching the independently fitted spectral radius
$\rho(\Bmat)=\rhoB$ to three decimal places. We emphasize what this rules out: had the cascade
been governed by the operator norm $\sigma_{\max}(\Bmat)=\sigmaxB$, leakage would have grown
tenfold per pass; had the fit been spurious, there is no reason a regression on single swaps
should predict the fifth-pass decay rate of an entirely separate experiment to three decimals.
The caveat: the ratio sequence is not perfectly monotone ($0.915\to0.893$), so the idealized
log-convex two-rate curve is violated by one small wobble within chain-to-chain noise.

\begin{figure}[t]
\centering
\includegraphics[width=0.85\columnwidth]{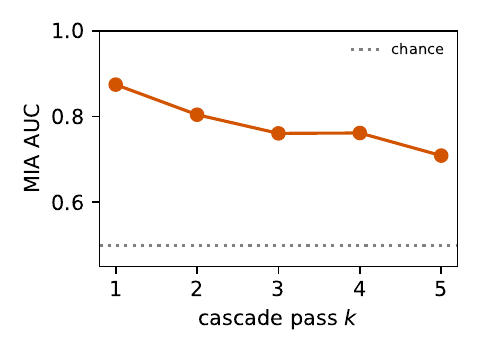}
\caption{Membership-inference AUC versus cascade depth (\Nchains{} chains). The attack decays
toward chance as Prediction~\ref{pred:floor} requires, but remains at $\mauclo$ after five
passes because the tail rate $\rho(\Bmat)$ is close to 1.}
\label{fig:miacascade}
\end{figure}

\smallskip
\noindent\textbf{Prediction~\ref{pred:floor} (floor): consistent, not yet reached.} The
non-member scores across passes have overall median $\mfloor$, in good agreement with the
Lyapunov-equation plateau $0.0055$ computed from the fit alone. Measured leakage at pass 5
($\mleakfive$) is still roughly $7\times$ the floor; extrapolating the confirmed tail rate, the
excess would not drop below $0.01$ until pass ${\sim}\passestofloor$. The membership attack
weakens with depth exactly as predicted --- AUC $\maucone\to0.804\to0.761\to0.762\to\mauclo$
(Figure~\ref{fig:miacascade}) --- but after five swaps of every image, the adversary of Part~1
still achieves AUC $\mauclo$. Dilution moves the needle; it does not, at practical depths, reach
chance.

\smallskip
\noindent\textbf{Prediction~\ref{pred:firstpass} (first pass dominates): confirmed.} The first
pass removes $0.525$ of absolute gallery similarity; all four subsequent passes combined remove
$0.044$ --- a $12{:}1$ asymmetry in favor of the first swap.

\smallskip
\noindent\textbf{Prediction~\ref{pred:persistent} (persistent directions): weak support only.}
Chains whose targets align with the top eigenspace of $\Bmat$ do leak more at pass 1 (Spearman
$\rho_s=0.21$) and survive marginally longer ($\rho_s=0.11$), with the predicted sign but small
effect sizes. We report this prediction as \emph{not convincingly confirmed}. In hindsight this
is expected: with $\rho(\Bmat)$ eigenvalues spread densely below $0.9$
(Figure~\ref{fig:spectrum}) and large stochastic innovations each pass, alignment with the
top eigenvectors is a noisy predictor of individual-chain fate at $k\leq5$.

\smallskip
\noindent\textbf{Where the model fails: levels.} The Monte Carlo rollout over-predicts pass-1
leakage ($0.168$ vs $0.091$) and then under-predicts the tail ($0.018$ vs $0.047$ at pass 5) ---
the measured curve is flatter than the simulated one even though their late \emph{rates} agree.
The discrepancy has a mechanical explanation: the operator was fitted on swaps of \emph{natural}
photographs, but from pass 2 onward the cascade feeds the swapper its own outputs, whose
embeddings lie off the natural image manifold. Consistent with this, residual--state
correlations in the cascade ($0.04$--$0.06$, Appendix~\ref{sec:model-adequacy}) exceed their
single-swap counterparts ($\approx0.01$--$0.02$): on its own outputs the swapper deviates from
the fitted mean dynamics in a way that mildly favors preserving the existing face. The honest
summary: the model identifies the decay \emph{mechanism} and \emph{rate} correctly, but
extrapolating absolute levels across this domain shift requires conservatism --- in our data, real
cascades erase \emph{more slowly} than the naive simulation, never faster. For a privacy
mechanism the optimistic direction is the dangerous one, which is precisely why we verify on
real chains.

\subsection{Generality: BlendFace and CanonSwap}
\label{sec:verify-cross}
\begin{figure}[t]
\centering
\includegraphics[width=\columnwidth]{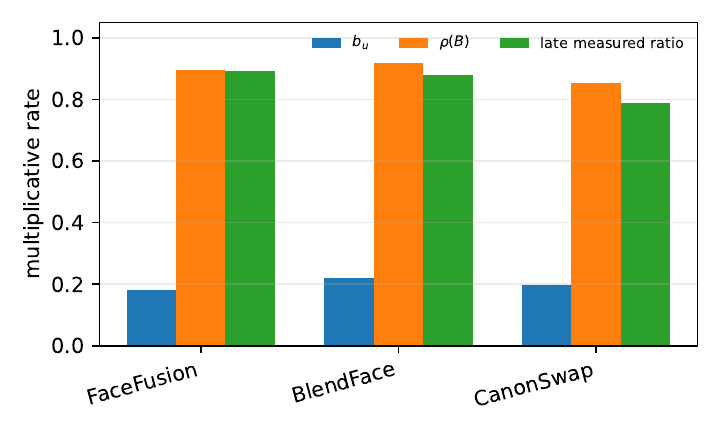}
\caption{Cross-tool rates: first-pass in-direction gain $b_u$, fitted spectral radius
$\rho(\Bmat)$, and measured late cascade ratio for the three modeled tools. The spectral
ordering (BlendFace slowest, CanonSwap fastest) matches the measured three-pass dilution
behavior, supporting Prediction~\ref{pred:crosstool}.}
\label{fig:crossrates}
\end{figure}

Prediction~\ref{pred:crosstool} concerns tool ordering, and here three-pass
dilution corpora (roughly 1{,}000 chains per tool, same VGGFace2 protocol, fresh donors per
pass) provide an independent test. Measured median-gallery leakage decays
$0.137\to0.104\to0.091$ for BlendFace (pass-to-pass ratios $0.76,\,0.88$) and
$0.124\to0.078\to0.062$ for CanonSwap (ratios $0.63,\,0.79$); per-chain decreases are highly
significant (Wilcoxon $p<10^{-46}$ for every consecutive pair, Friedman
$p<10^{-59}$). FaceFusion's measured ratios at the same depths are $0.677,\,0.873$. Levels and
late ratios line up with the independently fitted spectra
(Figure~\ref{fig:crossrates}): BlendFace, with the largest $\rho(\Bmat)=\bfRhoB$, has the
slowest tail ($\approx0.88$ by pass 3 and still climbing); CanonSwap, with the smallest
($\csRhoB$), dilutes fastest ($\approx0.79$); FaceFusion sits between, exactly as the
spectral ordering requires. Three passes are too shallow to pin each tool's asymptotic ratio, so
we read this as confirming the \emph{ordering} rather than exact tail
values. Notably, the ordering is not implied by single-swap leakage: BlendFace and
CanonSwap start at nearly the same pass-1 leakage ($0.137$ vs $0.124$), yet their cascades
separate with depth exactly as the spectra predict --- by pass 3, CanonSwap has shed
nearly twice the fraction of its initial leakage. The spectral radius captures something
no single-pass measurement reveals.

\section{Discussion: Privacy Implications}
\label{sec:discussion}

\noindent\textbf{Single-swap anonymization should be presumed leaky.} A face-swap that looks
right and passes a donor-match audit can
still expose $30$--$61\%$ of protected individuals to a confident membership adversary (TPR at
$1\%$ FPR, Table~\ref{tab:mia-metrics}). Evaluations of swap-based anonymizers should therefore
report TPR at low FPR against an explicitly measured non-member control population --- and
closed-set CMC where the candidate pool is enumerable --- not only mean match-score suppression.
The non-member baseline defines what ``anonymized'' can even mean: no mechanism that outputs a
face can push a target's similarity below that of a random stranger.

\smallskip
\noindent\textbf{Tool choice dominates at one pass; spectra govern at depth.} Across the five
credible anonymizers, single-pass leakage spans $0.048$--$0.137$ ($6\times$ in TPR at $1\%$
FPR), so a data owner constrained to one pass should prefer tools that synthesize farther from
both inputs (E4S in our cohort), accepting weaker donor resemblance. With multiple passes the
relevant property changes: the first pass is governed by $b_u$, which is similar across tools
($0.18$--$0.22$), while later passes are governed by $\rho(\Bmat)$, which differs meaningfully
($0.85$--$0.92$) and is invisible to single-pass benchmarking --- BlendFace and CanonSwap start
at near-identical pass-1 leakage yet separate steadily with depth. Spectral fitting is, to our
knowledge, the only practical way to anticipate this short of running deep cascades per tool.

\smallskip
\noindent\textbf{Dilution works, slowly, with quantifiable diminishing returns.}
Operationally, the first swap buys an
$\sim85\%$ leakage reduction; each later swap removes only $\sim11\%$ of what remains
(rate $\approx0.89$ for FaceFusion). Driving excess leakage below $0.01$ --- roughly where the
MIA advantage becomes marginal --- extrapolates to $\passestofloor$ passes, each costing
generation time and visual degradation; and since the measured tail runs \emph{above} the model
rollout, simulation-based dilution budgets are lower bounds. The compensating benefit is an
interpretable, checkable guarantee: because the floor is the non-member baseline and the rate is
measurable, a data owner can state ``after $k$ passes, expected excess leakage is at most
$(\text{pass-1 excess})\cdot\rho^{k-1}$'' instead of the purely observational ``we ran the tool
and scores looked low.''

\smallskip
\noindent\textbf{The spectral radius is a design target.} $\rho(\Bmat)$ and $b_u$ are, in
principle, differentiable functionals of the swapping network, estimable from a few thousand
swaps (Appendix~\ref{sec:model-corpus}). Training a swapper with an explicit penalty on
target-transfer --- minimizing the empirical Rayleigh gain of the induced $\Bmat$, or
regularizing its spectrum --- would attack the leakage channel at its source, and the same
machinery gives third parties a certification protocol: fit the operator on a probe corpus,
report $(b_u,\rho(\Bmat))$ and the implied dilution schedule.

\smallskip
\noindent\textbf{Why does the channel exist at all?} Swappers preserve pose, expression,
lighting, and face shape from the target by design; recognition embeddings are trained to be
invariant to exactly these factors, but the invariance is imperfect, and residual covariance
between ``attributes'' and identity lets preserved attributes drag identity along. The fitted
$\Bmat$ quantifies the aggregate strength of that pathway ($b_u\approx0.2$: about a fifth of the
target's identity direction survives one swap) without attributing it to specific attributes;
decomposing $\Bmat$ against interpretable factors --- geometry, texture, blending mask --- is a
natural next step that the operator framework makes well-posed.

\section{Limitations}
\label{sec:limitations}
\noindent\textbf{Scope of the adversary.} Our attacks threshold or rank cosine similarities from
public embeddings. Stronger adversaries --- tool-aware, trained on swap outputs, or combining
multiple released images of the same target --- would extract more; our leakage numbers are lower
bounds. Conversely, we study identity inference only; attribute inference from preserved
non-facial context is a separate channel we do not measure.

\smallskip
\noindent\textbf{Model fidelity.} The affine model explains $59\%$ of held-out swap-embedding
variance for FaceFusion but only $40\%$ and $34\%$ for BlendFace and CanonSwap; for the latter
two, spectral conclusions rest on a coarser approximation. Residual anisotropy and
non-Gaussianity mean the model should not be used for fine-grained distributional claims, and
the deep-cascade level mismatch shows that off-manifold extrapolation is optimistic. We have
been careful to claim only what survived verification: mechanisms, rates, floors, orderings.

\smallskip
\noindent\textbf{Coverage.} Reliable operator spectra require $10^3$--$10^4$ swaps per tool
(Figure~\ref{fig:learning}); we therefore fit three of the five donor-retaining tools, and the
five-pass cascade exists for FaceFusion only (BlendFace and CanonSwap have three passes).
DiffFace and E4S have single-pass leakage measurements but no fitted operators. Extending the
spectral analysis to them --- particularly E4S, the most private single-pass tool --- is mechanical
but not free: at the single-swap generation rates measured in our runs ($\approx$31\,s per E4S
swap and $\approx$150\,s per DiffFace swap), a 24k-swap operator corpus would cost roughly 210
and 1{,}000 GPU-hours, respectively.

\smallskip
\noindent\textbf{Data and embeddings.} All experiments use VGGFace2 (celebrity faces,
GFPGAN-normalized); demographic structure beyond the MAAD-matched control of
Section~\ref{sec:part1-robust} is unexplored, and clinical or surveillance imagery may behave
differently. Two recognition embeddings anchor the measurements; a future, stronger recognizer
would see more leakage (the Facenet-vs-ArcFace gap in
Table~\ref{tab:mia-metrics} already shows recognizer strength translating into attack strength).

\smallskip
\noindent\textbf{Utility.} We quantify privacy, not the utility cost of cascading. Repeated
swapping visibly degrades fine image detail and progressively replaces the donor identity of
earlier passes with that of the final donor; a full privacy--utility frontier is future work.

\section{Conclusion}
\label{sec:conclusion}
Face-swapping anonymization rests on a plausible but, as we show, incomplete security argument.
Across seven modern tools, a single swap leaves target identity in the released image at levels
an off-the-shelf thresholding adversary can exploit --- up to $61\%$ of targets recovered at
$1\%$ false-positive rate among tools that genuinely transfer the donor. Modeling the swapper as
an affine stochastic operator on the adversary's embedding space turns this observation into a
mechanism: the target enters a cascade once and is propagated by the target-transfer operator
$\Bmat$, whose Rayleigh gain explains the large first-pass suppression, whose spectral radius
dictates the slow geometric tail, and whose stability fixes the non-member baseline as the
privacy floor. The model's predictions were tested on dedicated multi-pass dilution data: the
two-rate decay and the spectral tail rate were confirmed to three decimal places for FaceFusion
and ordered three tools correctly, while absolute deep-cascade levels exposed the model's
optimism under domain shift --- a failure we report with the same prominence as the successes,
because for privacy mechanisms the direction of model error matters.

The broader contribution is methodological. Privacy claims for generative anonymization are
today mostly observational: run the tool, measure the scores. The operator framework makes such
claims \emph{explainable} --- leakage is located in a measurable object with
interpretable spectra --- and \emph{perfectible}: the spectrum is a concrete training and
certification target for the next generation of face-swapping anonymizers. Until such tools
exist, our results counsel caution: a swapped face is anonymous only up to an operator that
current systems leave far from zero.

\bibliographystyle{IEEEtran}
\bibliography{refs}

\appendices

\section{Example Swap Outputs}
\label{app:examples}
\begin{figure*}[!t]
\centering
\includegraphics[width=0.92\textwidth]{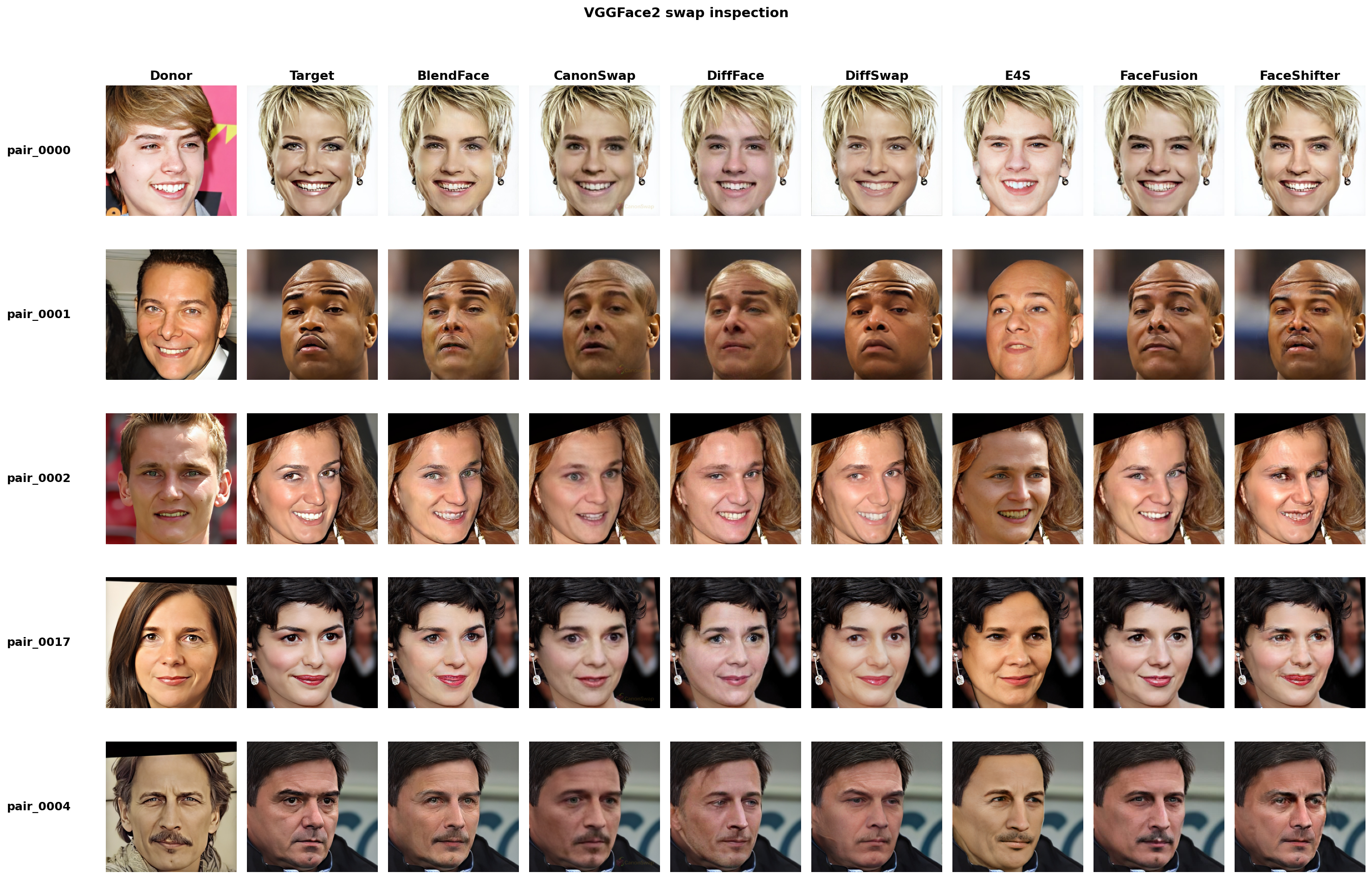}
\caption{Face-swapping in practice: five donor--target pairs from our VGGFace2 evaluation
(rows) swapped by each of the seven tools studied (columns~3--9). Each output preserves the
target image's pose, hair, background, and expression while the inner face is replaced. Visually,
the outputs are donor-like; the question this paper asks is how much \emph{target} identity
nevertheless survives.}
\label{fig:swapexamples}
\end{figure*}

Figure~\ref{fig:swapexamples} shows example outputs of the seven face-swapping tools evaluated in
Part~1, on five donor--target pairs from the benchmark of Section~\ref{sec:part1-method}.

\section{Identity Embedding Clusters}
\label{app:embclusters}
\begin{figure}[t]
\centering
\includegraphics[width=0.8\columnwidth]{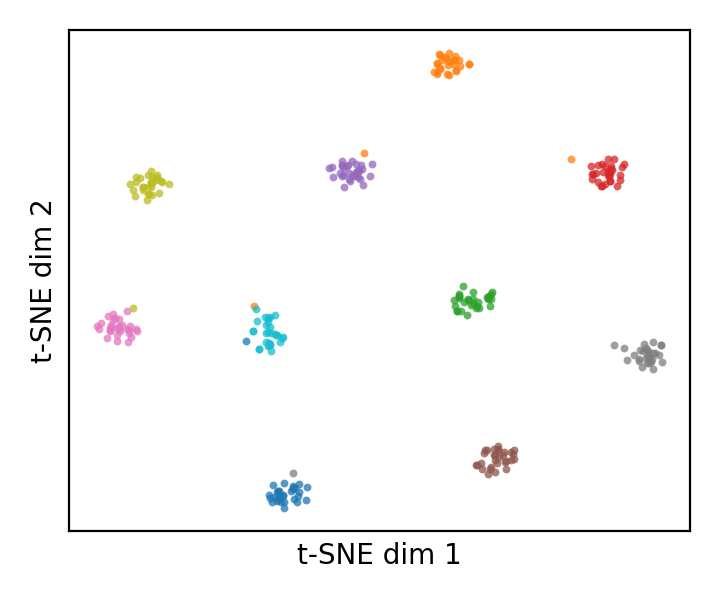}
\caption{Identity embeddings of ten randomly chosen VGGFace2 subjects (up to 30 images each,
InsightFace \texttt{buffalo\_l}), visualized with t-SNE. Images of the same person form tight,
well-separated clusters; cosine similarity in this space is therefore a meaningful identity
signal.}
\label{fig:embclusters}
\end{figure}

Figure~\ref{fig:embclusters} visualizes the identity-clustering property of the embedding space
described in Section~\ref{sec:prelim-extract}.

\section{Closed-Set Re-Identification Results}
\label{app:reid}
\begin{figure*}[!t]
\centering
\includegraphics[width=0.88\textwidth]{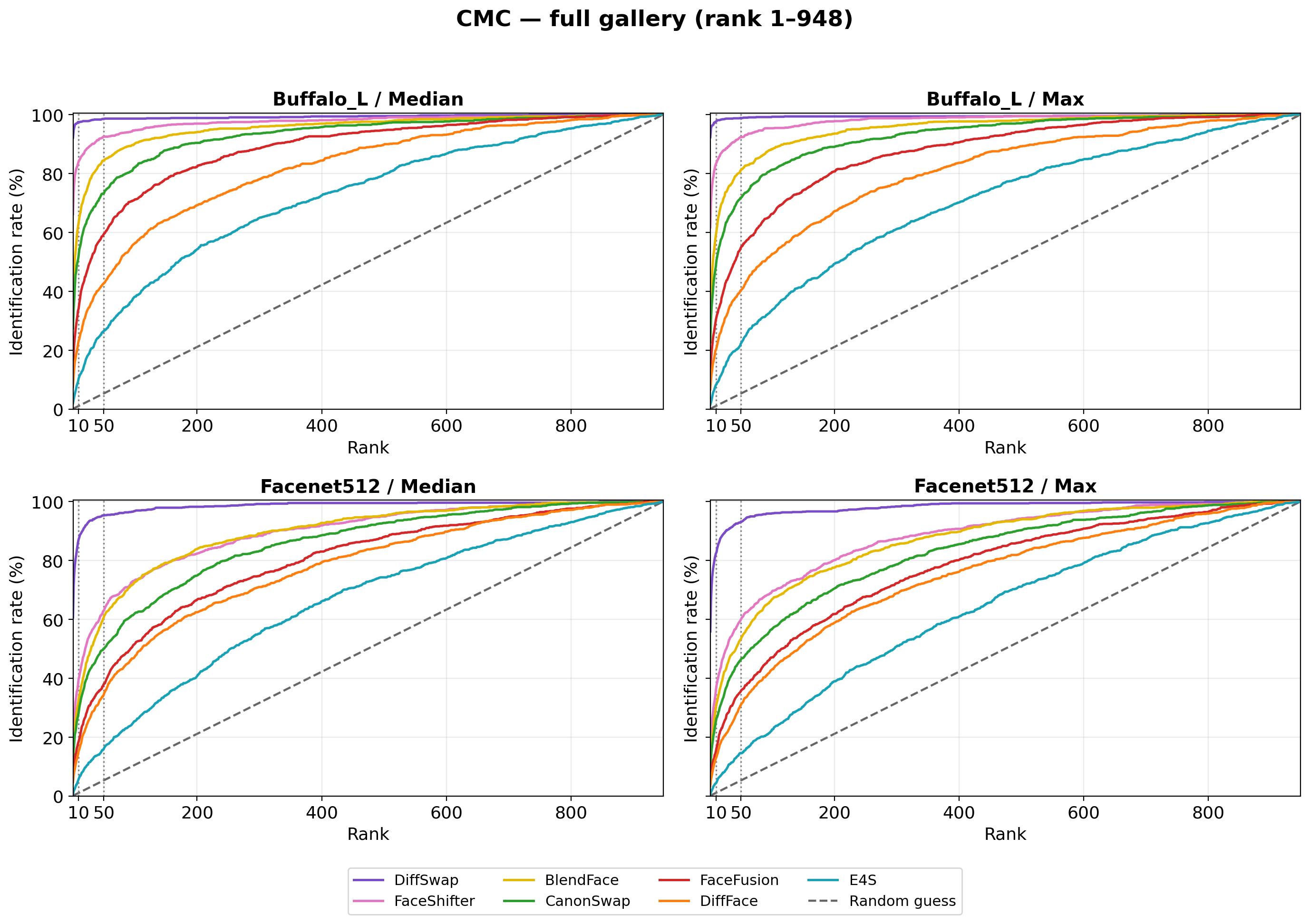}
\caption{Closed-set re-identification CMC curves against a full gallery of 948 candidate
identities (top: \texttt{buffalo\_l}; bottom: Facenet512; left: median aggregation, right: max).
Random guessing is the diagonal. Every tool lies far above chance: even for the most private
tool (E4S), the true target is ranked in the top 50 of 948 candidates for one quarter of swaps
under \texttt{buffalo\_l}.}
\label{fig:cmc}
\end{figure*}

The membership attack of Section~\ref{sec:part1-mia} asks ``Is this person in the release?''. A
complementary adversary already knows the release contains one of $N$ known candidates and asks
``Which one?''. We evaluate this closed-set re-identification problem on the same benchmark: for
each swap, all 948 candidate identities (the true target plus 947 decoys) are ranked by
subject-level similarity and the rank of the true target is recorded.
Figure~\ref{fig:cmc} shows the resulting CMC curves; Table~\ref{tab:cmc} summarizes rank-$k$
identification rates.

\begin{table}[H]
\centering
\caption{Closed-set identification rates (\%) at rank $k$ out of 948 candidates
(VGGFace2, \texttt{buffalo\_l}, median aggregation). Random guessing: rank-1 $=0.1\%$,
rank-50 $=5.3\%$.}
\label{tab:cmc}
\begin{tabular}{lcccc}
\toprule
Tool & Rank-1 & Rank-5 & Rank-10 & Rank-50 \\
\midrule
FaceFusion  & 13.7 & 26.8 & 34.1 & 59.3 \\
DiffFace    &  7.6 & 17.2 & 23.2 & 42.7 \\
BlendFace   & 34.2 & 53.3 & 63.3 & 84.4 \\
E4S         &  \textbf{2.0} &  \textbf{6.1} & \textbf{10.3} & \textbf{26.5} \\
CanonSwap   & 27.5 & 45.5 & 52.1 & 73.3 \\
\midrule
FaceShifter & 66.2 & 80.5 & 84.1 & 92.5 \\
DiffSwap    & \underline{92.2} & \underline{96.8} & \underline{97.5} & \underline{98.5} \\
\bottomrule
\end{tabular}
\end{table}

For BlendFace, the single most
similar identity among 948 candidates is the true target in $34\%$ of cases --- 340 times the
chance rate --- and the target is in the top ten for $63\%$ of swaps. Even E4S, whose membership
ROC looked comparatively benign, places the true target in the top 50 ($5\%$ of the gallery) for
$26.5\%$ of swaps, five times chance. Re-identification compounds leakage: the adversary does not
need a high absolute score, only a score higher than 947 unrelated identities, and the
target-specific signal documented in Part~1 is exactly what wins that comparison.

\section{Is the Affine Model Adequate?}
\label{sec:model-adequacy}
This appendix details the five adequacy checks summarized in Section~\ref{sec:model-fit},
reporting failures as well as successes.

\smallskip
\noindent\textbf{(1) It beats every simpler baseline by a wide margin.}
Table~\ref{tab:ladder} compares, on the identity-disjoint FaceFusion test set, a ladder of
predictors of the swap embedding. Predicting the global mean captures nothing
($0.122$); copying the donor embedding --- the model implicitly assumed by ``the output is
the donor'' --- reaches $0.654$; an affine transform of the donor alone reaches $0.722$; and
adding the target term lifts held-out cosine to
$\fitcos$ with $R^2=\fitRtwo$. The gap between the last two rows is the statistical signature of
the leakage channel: target information improves prediction beyond anything the donor provides,
by far too much to be noise on $\ffTest{}$ test
triplets. Conversely, a target-only model is poor ($0.282$), confirming the output is primarily
donor, and a $k$-nearest-neighbor regressor in the concatenated input space
($k=25$) --- a nonparametric stand-in for ``some smooth nonlinear map'' --- achieves only
$0.442$ at this corpus size, far below the affine model.

\begin{table}[t]
\centering
\caption{Predicting held-out FaceFusion swap embeddings (\ffTest{} identity-disjoint test
triplets, raw ArcFace space): a ladder of baselines. The donor$+$target affine model
\eqref{eq:swap} is the simplest predictor that captures the leakage channel.}
\label{tab:ladder}
\begin{tabular}{lcc}
\toprule
Predictor & Mean $\cos(\hat{\bm{s}},\bm{s})$ & $R^2$ \\
\midrule
Global mean embedding        & 0.122 & 0.00 \\
Copy target ($\hat{\bm{s}}=\bm{t}$)   & 0.175 & $-0.54$ \\
Target-only ridge            & 0.282 & 0.07 \\
Copy donor ($\hat{\bm{s}}=\bm{d}$)    & 0.654 & 0.35 \\
Donor-only ridge             & 0.722 & 0.51 \\
$k$NN ($k{=}25$, donor$\oplus$target) & 0.442 & 0.16 \\
\textbf{Affine donor$+$target \eqref{eq:swap}} & \textbf{0.769} & \textbf{0.585} \\
\bottomrule
\end{tabular}
\end{table}

\smallskip
\noindent\textbf{(2) Explicit nonlinear corrections add essentially nothing.} We augmented the
affine model with batteries of nonlinear features of $(\bm{d},\bm{t})$ --- elementwise products
and squares, donor--target interaction terms, and norm/angle features --- and refit on the same
splits. The best feature union improves held-out $R^2$ by about $0.001$. At this corpus size,
whatever structure the affine model misses is not a smooth low-order function of the inputs; it
behaves like noise, which is precisely how \eqref{eq:swap} treats it.

\begin{figure}[!htb]
\centering
\includegraphics[width=0.85\columnwidth]{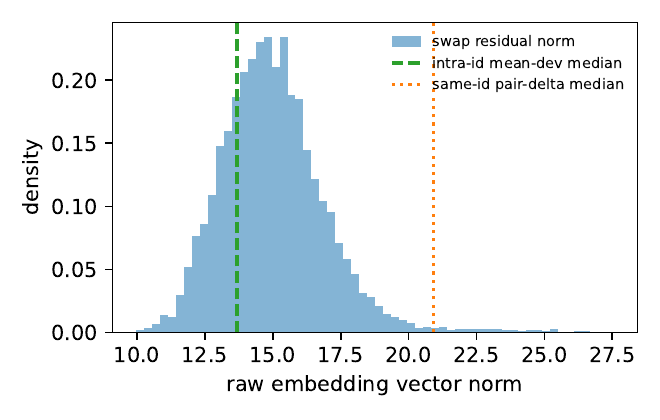}
\caption{Scale of the affine-model residual compared with natural intra-identity embedding
variation in VGGFace2. The residual is large (the model is approximate) but no larger than the
ordinary spread of same-person embeddings, supporting its treatment as stochastic innovation.}
\label{fig:modelval}
\end{figure}

\smallskip
\noindent\textbf{(3) The residual has the scale of natural identity noise.} The model leaves
$\stochfrac$ of swap-embedding variance unexplained for FaceFusion --- a substantial residual.
Its scale, however, is informative: the median residual
norm is $\modelresidmed$, close to the median deviation of ordinary same-person
VGGFace2 embeddings from their identity mean ($\intraidmed$ over 3{,}000 identities). The affine
model thus predicts the swap embedding about as accurately as one photograph of a person
``predicts'' another photograph of the same person, so the residual is plausibly the same kind
of variation --- pose, expression, illumination, capture noise --- plus generator-specific
randomness, rather than a missing deterministic signal (Figure~\ref{fig:modelval}).

\smallskip
\noindent\textbf{(4) The residual does not hide the target.} The cascade analysis
requires one property above all: the residual must not systematically re-inject the protected
identity. We test this directly on held-out data. The mean cosine between residual directions
and the corresponding target direction is $-0.002$ (mean absolute per-dimension correlation
$0.009$); against the propagated component $\Bmat\bm{t}$ the values are $-0.025$ and $0.023$.
For comparison, the same statistics against the donor term are equally small. In the deep
cascade of Section~\ref{sec:verify}, pass-wise mean absolute correlations between residuals and
the propagated state remain $0.04$--$0.06$ --- larger than in single swaps (an early sign of the
domain shift discussed there) but still small in absolute terms. Residuals may be big, but they
are, to good approximation, \emph{blind to the target}.

\smallskip
\noindent\textbf{(5) Distributional assumptions: what holds and what does not.} We subjected
the FaceFusion residuals to a battery of formal tests, and report the verdicts without
varnish. \emph{Unbiasedness holds approximately}: the residual mean norm is about $1.3\%$ of the
mean residual norm --- negligible in magnitude, though statistically distinguishable from zero at
$N{=}10{,}667$. \emph{Exogeneity holds}: neither linear nor $k$NN regressors can predict the
residual from $(\bm{d},\bm{t})$ (out-of-sample $R^2<0$); there is no recoverable leftover signal
in the inputs. \emph{Homoscedasticity holds only partially}: residual magnitude correlates
mildly with how strongly the donor was written ($r\approx-0.23$) --- swaps that transfer the donor
well are slightly more predictable. \emph{Isotropy fails}: the residual covariance is strongly
anisotropic, concentrated in a few hundred directions. \emph{Gaussianity fails}: residuals are
heavier-tailed than Gaussian. Geometrically, residuals live almost entirely in the tangent space
of the embedding sphere ($\approx98\%$ of energy), consistent with the constant-norm observation
of Section~\ref{sec:model-fit}.

These verdicts matter for scope. The cascade predictions of Section~\ref{sec:cascade} concern
\emph{first moments and decay rates}: they require the mean dynamics \eqref{eq:swap}, residual
unbiasedness, and residual--target uncorrelatedness --- all supported. They do not require
isotropy or Gaussianity, which would only be needed for finer distributional predictions (e.g.\
exact per-pass score histograms). Indeed, when we push model samples through the Part~1 scoring
pipeline with a heteroscedastic noise model, the predicted target/non-member/donor score
ordering and magnitudes are reproduced only approximately (donor scores are underestimated),
which is why we lean on the model for rates and floors, not for exact distributions.

\section{How Much Data Does Operator Identification Need?}
\label{sec:model-corpus}
\begin{figure}[H]
\centering
\includegraphics[width=0.85\columnwidth]{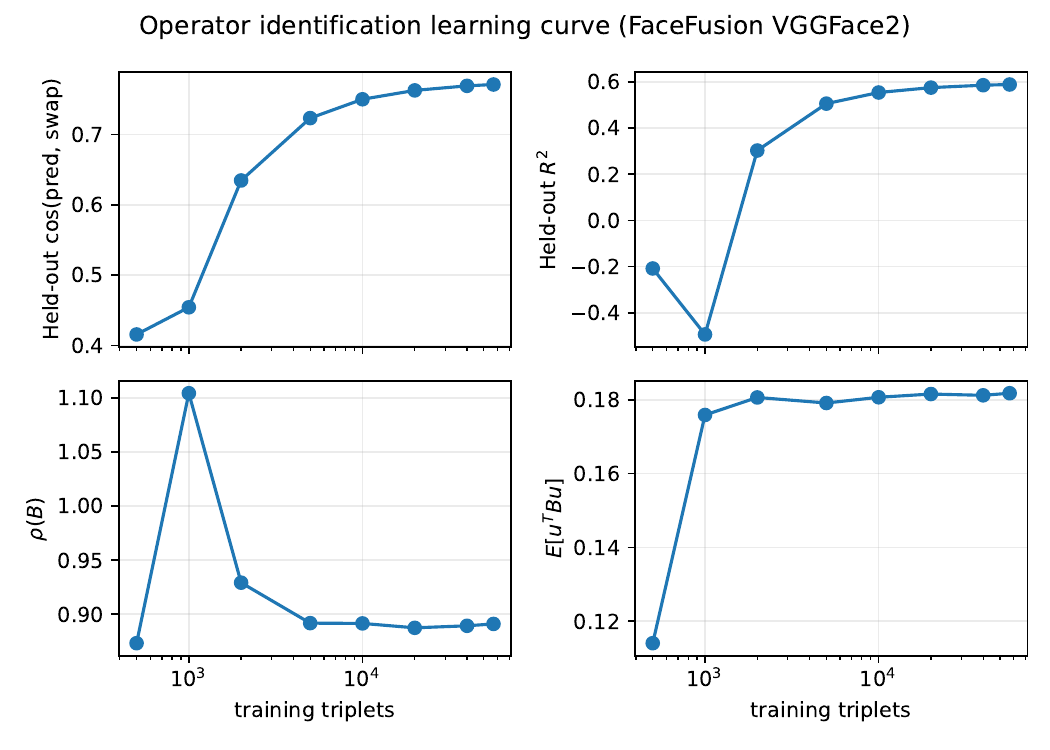}
\caption{Operator identification versus training-corpus size (FaceFusion, fixed identity-disjoint
test set). Small corpora yield unstable spectra --- at $N{=}1{,}000$ the estimated $\rho(\Bmat)$
exceeds 1 --- while $\rho(\Bmat)$ stabilizes near \rhoStableN{} triplets and held-out fit quality
near \fitStableN{}.}
\label{fig:learning}
\end{figure}

Estimating two $512\times512$ operators is a high-dimensional regression, and the spectral
quantities we care about are notoriously sensitive to undersampling. Figure~\ref{fig:learning}
traces fit quality and spectral estimates as the training corpus grows, with the test set held
fixed. The behavior is cautionary: at 500--1{,}000 training triplets the fits are unusable ---
held-out $R^2$ is negative and the estimated spectral radius even exceeds $1$ at $N{=}1{,}000$,
which would (wrongly) predict that cascades \emph{amplify} the target. The spectral radius
stabilizes to within $1\%$ of its final value by roughly \rhoStableN{} triplets; held-out cosine
and $R^2$ converge more slowly, by roughly \fitStableN{}. Two consequences follow. First, the
\npairs-pair single-swap corpora of Part~1 are an order of magnitude too small for spectral
fitting, which
is why Part~2 required dedicated corpora and why we fit three tools rather than seven. Second,
the in-direction gain $b_u$ stabilizes much earlier (by ${\sim}1{,}000$ triplets), so first-pass
leakage is cheap to estimate even when the full spectrum is not.

\section{Visual Quality Under Dilution}
\label{app:dilutionsamples}
\begin{figure}[H]
\centering
\includegraphics[width=0.95\columnwidth]{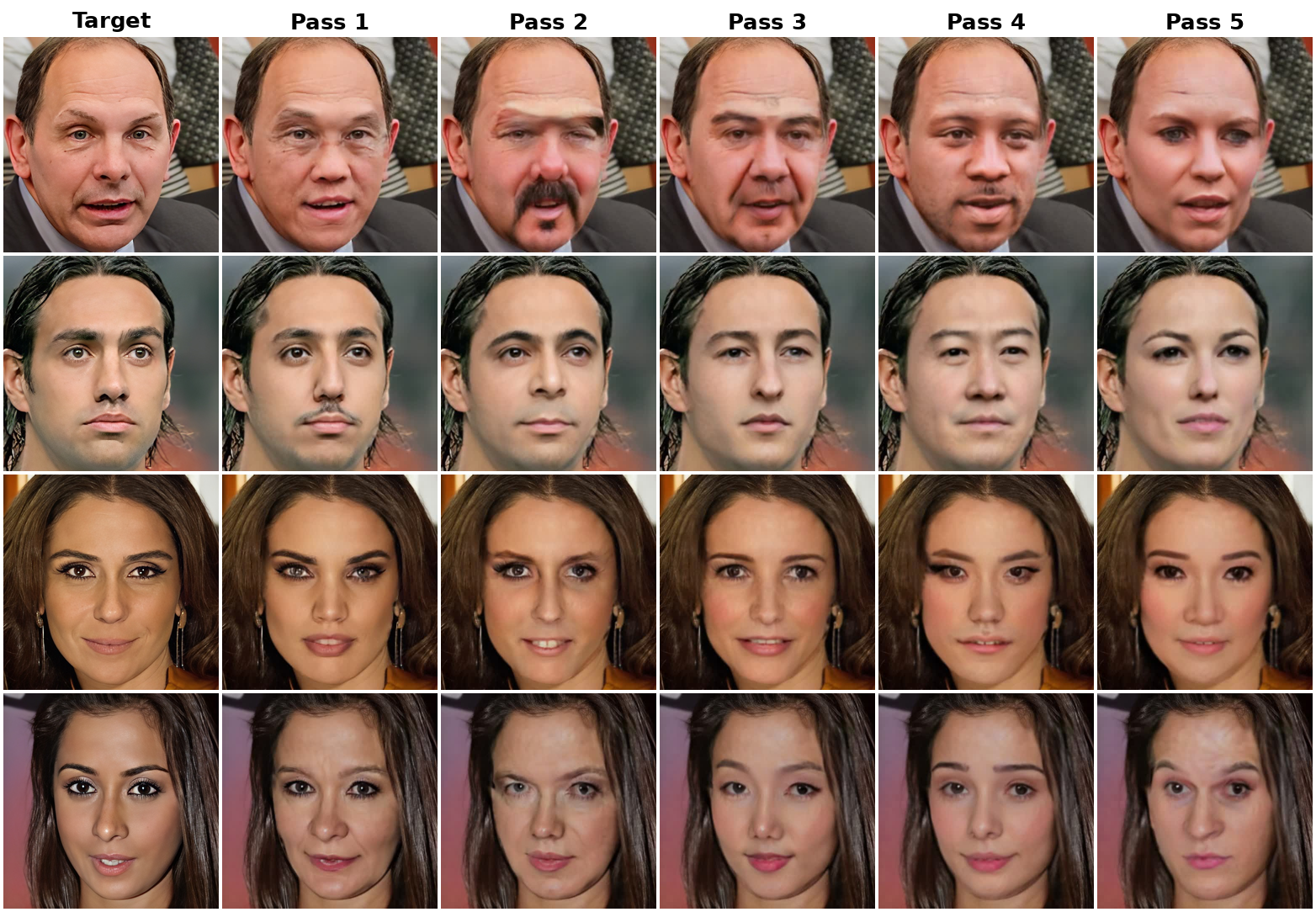}
\caption{Dilution cascades in image space: four chains (rows) from the five-pass FaceFusion
cascade. The leftmost column is the original target image; column $k$ shows the output after
$k$ passes, each with a fresh donor. Outputs remain photorealistic and retain the target
image's pose, expression, and context throughout, with only mild loss of fine detail.}
\label{fig:dilutionsamples}
\end{figure}

Figure~\ref{fig:dilutionsamples} shows example chains from the five-pass FaceFusion cascade of
Section~\ref{sec:verify-protocol}. Photorealism is largely preserved across passes: pose,
expression, hair, and background remain those of the original target image, while the inner
face drifts toward each pass's fresh donor. Only mild smoothing of fine detail accumulates with
depth, so the privacy gains of dilution are not paid for with obvious visual artifacts.

\end{document}